\PassOptionsToPackage{table}{xcolor}
\documentclass[sigconf,10pt,screen]{acmart}
\AtBeginDocument{%
  }

\copyrightyear{2025}
\acmYear{2025}
\setcopyright{cc}
\setcctype{by-nc}
\acmConference[ICS '25]{2025 International Conference on Supercomputing}{June 8--11, 2025}{Salt Lake City, UT, USA}
\acmBooktitle{2025 International Conference on Supercomputing (ICS '25), June 8--11, 2025, Salt Lake City, UT, USA}
\acmDOI{10.1145/3721145.3725763}
\acmISBN{979-8-4007-1537-2/2025/06}

\newcommand{\proj}{{\tt NeurLZ}}
\newcommand{\sproj}{{SFLZ}}

\usepackage{color}
\usepackage{enumitem}
\usepackage{tikz}
\usetikzlibrary{positioning}
\usepackage{svg}
\usepackage{soul}
\usepackage{multirow} %
\usepackage{url} %
\usepackage[misc]{ifsym}

\usepackage{tikz}
\newcommand*\Circled[2][gray!40]{%
	\tikz[baseline=(char.base)]{\node[
        shape=circle, draw=none,  thick, 
        fill=#1 ,inner sep=0.9pt] (char) 
    {\textcolor{black}{#2}}; 
}}

\usepackage{framed}

\definecolor{that-yellow}{HTML}{ddc26f}

\newenvironment{formal}{%
  \MakeFramed{\advance\hsize-\width\FrameRestore}%
  \noindent\hspace{-4.pt}%
}
{%
  \endMakeFramed%
}

\newcommand{\FakeParagraph}[1]{\vspace{0.1cm}\noindent\textbf{$\blacksquare$ #1.}\hspace{1em}} 

\begin{document}

\title{\proj: An Online Neural Learning-Based Method to Enhance Scientific Lossy Compression}

\settopmatter{authorsperrow=4}

\author{\href{https://orcid.org/0009-0007-9473-6703}{Wenqi Jia}}
\affiliation{%
  \institution{\normalsize UT Arlington}
  \country{\normalsize Arlington, TX, USA}
}

\author{\href{https://orcid.org/0009-0004-3516-0431}{Zhewen Hu}}
\affiliation{%
  \institution{\normalsize \makebox[8em][c]{Texas A\&M University}}
  \country{\normalsize College Station, TX, USA}
}

\author{\href{https://orcid.org/0009-0002-6353-3659}{Youyuan Liu}}
\affiliation{%
  \institution{\normalsize Temple University}
  \country{\normalsize Philadelphia, PA, USA}
}

\author{\href{https://orcid.org/0009-0003-8937-4067}{Boyuan Zhang}}
\affiliation{%
  \institution{\normalsize Indiana University}
  \country{\normalsize Bloomington, IN, USA}
}

\author{\href{https://orcid.org/0000-0001-6317-2940}{Jinzhen Wang}}
\affiliation{%
  \institution{\normalsize UNC Charlotte}
  \country{\normalsize Charlotte, NC, USA}
}

\author{\href{https://orcid.org/0000-0003-0177-502X}{Jinyang Liu}}
\affiliation{%
  \institution{\normalsize \makebox[8em][c]{University of Houston}}
  \country{\normalsize Houston, TX, USA}
}

\author{\href{https://orcid.org/0000-0002-2697-7042}{Wei Niu}}
\affiliation{%
  \institution{\normalsize \makebox[8em][c]{University of Georgia}}
  \country{\normalsize Athens, GA, USA}
}

\author{\href{https://orcid.org/0000-0002-7543-2479}{Stavros Kalafatis}}
\affiliation{%
  \institution{\normalsize \makebox[8em][c]{Texas A\&M University}}
  \country{\normalsize College Station, TX, USA}
}

\author{\href{https://orcid.org/0000-0002-9548-1227}{Junzhou Huang}}
\affiliation{%
  \institution{\normalsize UT Arlington}
  \country{\normalsize Arlington, TX, USA}
}

\author{\href{https://orcid.org/0009-0009-9250-0611}{Sian Jin}}
\affiliation{%
  \institution{\normalsize Temple University}
  \country{\normalsize Philadelphia, PA, USA}
}

\author{\href{https://orcid.org/0000-0002-4444-3634}{Daoce Wang*}}
\affiliation{%
  \institution{\normalsize Indiana University}
  \country{\normalsize Bloomington, IN, USA}
}

\author{\href{https://orcid.org/0000-0003-1101-9148}{Jiannan Tian*}}
\affiliation{%
  \institution{\normalsize \makebox[8em][c]{University of Kentucky}}
  \country{\normalsize Lexington, KY, USA}
}

\author{\href{https://orcid.org/0000-0002-5554-5417}{Miao Yin*$^\dagger$}}
\affiliation{%
  \institution{\normalsize UT Arlington}
  \country{\normalsize Arlington, TX, USA}
}

\renewcommand{\shortauthors}{Jia and Yin, et al.}

\begin{abstract}
  \let\thefootnote\relax\footnotetext{*Co-advising. $^\dagger$Corresponding author. Email: miao.yin@uta.edu.}Large-scale scientific simulations generate massive datasets, posing challenges for storage and I/O. Traditional lossy compression struggles to advance more in balancing compression ratio, data quality, and adaptability to diverse scientific data features. While deep learning-based solutions have been explored, their common practice of relying on large models and offline training limits adaptability to dynamic data characteristics and computational efficiency. To address these challenges, we propose \proj, a neural method designed to enhance lossy compression by integrating online learning, cross-field learning, and robust error regulation.
  Key innovations of {\proj} include: (1) {\it compression-time online neural learning} with lightweight skipping DNN models, adapting to residual errors without costly offline pertaining, (2) {\it the error-mitigating capability},  recovering fine details from compression errors overlooked by conventional compressors, (3) {\it $1\times$ and $2\times$ error-regulation modes}, ensuring strict adherence to $1\times$ user-input error bounds strictly or relaxed 2$\times$ bounds for better overall quality, and (4) {\it cross-field learning} leveraging inter-field correlations in scientific data to improve conventional methods. Comprehensive evaluations on representative HPC datasets, e.g., Nyx, Miranda, Hurricane, against state-of-the-art compressors show \proj's effectiveness. During the first five learning epochs, \proj~achieves an 89\% bit rate reduction, with further optimization yielding up to around 94\% reduction at equivalent distortion, significantly outperforming existing methods, demonstrating \proj's superior performance in enhancing scientific lossy compression as a scalable and efficient solution.
\end{abstract}




\begin{teaserfigure}
  \centering
  \begin{tikzpicture}
    \node{\includegraphics[width=\linewidth, trim={0 .3in 0 .4in}]{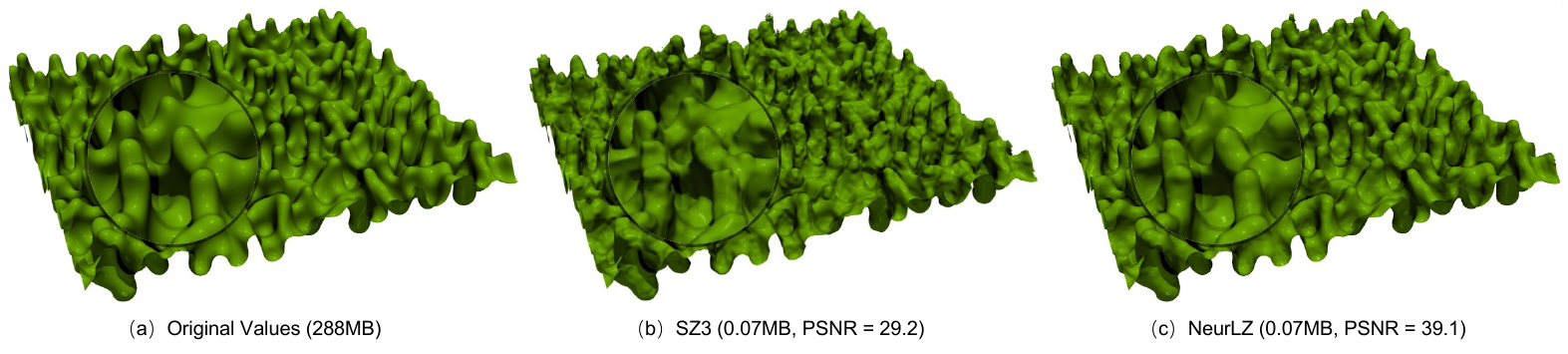}};
    \draw[ultra thick, black, opacity=.5] (-2.73in, .115in) circle (.39in);
    \draw[ultra thick, black, opacity=.5] (-.405in, .115in) circle (.39in);
    \draw[ultra thick, black, opacity=.5] (1.93in, .115in) circle (.39in);
    \draw[thick, white] (-2.73in, .115in) circle (.39in);
    \draw[thick, white] (-.405in, .115in) circle (.39in);
    \draw[thick, white] (1.93in, .115in) circle (.39in);
  \end{tikzpicture}
  \vspace{-5.5mm}
  \caption{Renderings of an isosurface from Miranda's Density field: (a) original values, (b) SZ3, and (c) \proj. The isosurface rendered by \proj~ exhibits noticeable visual improvements compared to that produced by SZ3.}
  \Description{Three isosurface renderings of Miranda's Density field are shown.}
  \label{fig:teaser}
\end{teaserfigure}

\maketitle

\section{Introduction}
The exponential growth in computational power has enabled complex scientific simulations, producing massive datasets across disciplines. For instance, the CESM climate simulation~\cite{cesm} generates terabytes of daily data volumes for post-processing~\cite{cesm_report}, highlighting the scale of data generated in modern scientific research. For warm data used in tasks like visualization and analysis~\cite{cappello2019use}, reducing storage overhead is crucial to minimize transmission delays and improve accessibility.
Further, for cold data stored long-term, minimizing storage overhead becomes a significant concern due to the substantial expenses involved, where real-time compression is unnecessary. Services like Amazon Glacier Deep Archive~\cite{amazon_s3_glacier} and Microsoft Azure Archive Storage~\cite{azure_archive_storage}, which use cost-effective storage mediums like magnetic tape~\cite{yoshida2020mass}, highlight the need for efficient compression solutions to balance accessibility, cost, and data fidelity~\cite{pernet2023long, mem2019tapes, Pelican, ECMWF}.

\begin{figure}
    \includegraphics[width=\linewidth, trim={0 0 0 .25in}]{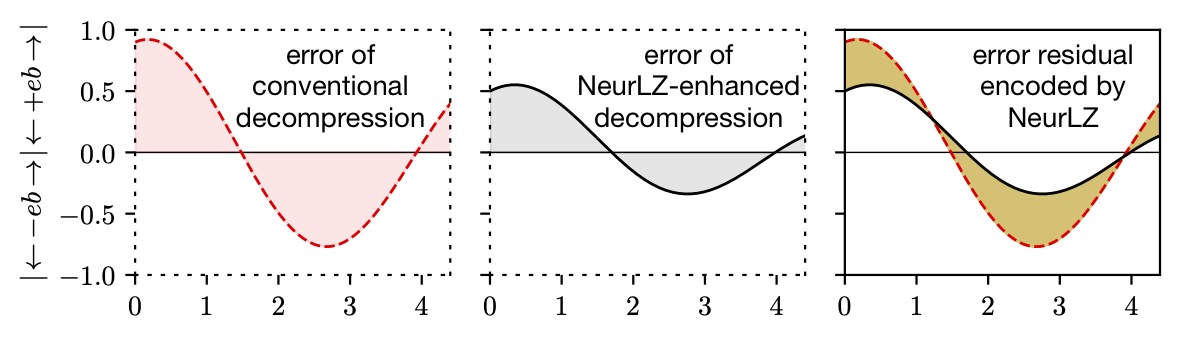}

    \footnotesize
    \resizebox{\linewidth}{!}{%
    \begin{tabular}{@{}c} file \\[-.6ex] format\end{tabular}~%
    \begin{tabular}{  | c | >{\cellcolor{that-yellow}}c | >{\cellcolor{that-yellow}}c |}
        \hline
        \begin{tabular}{@{}c@{}} compressed data from\\[-.6ex] conventional compressor \end{tabular}
         & \proj~model
         & \proj~outlier
        \\
        1.0667 MB (85.8\%)
         & 0.1089 MB (8.8\%)
         & 0.06739 MB (5.4\%)
        \\
        \hline
    \end{tabular}
    }
    \caption{(Top) compression error anatomy. (Bottom) the file format of {\proj} compression archive.}
    \label{fig:error-anatomy}
    \vspace{-\baselineskip}
\end{figure}

Initially, researchers utilized and developed lossless compression algorithms~\cite{ziv1977universal, GZIP1996, lindstrom2006fast, gailly2004zlib, collet2015zstandard} to mitigate data storage and transmission challenges. However, these techniques typically achieve only modest compression ratios at 1 to 3$\times$ when applied to scientific data~\cite{9378449}. Lossy compression algorithms~\cite{lindstrom2014fixed, 7516069, 10046076, 10.1145/3639259, 9458791, 7967203, liang2021mgard+, 10.1145/3295500.3356193, tao2019optimizing, 10.1145/3369583.3392688, 8622520, yang2024gpufastqlz, yang2024seqbench, liu2023faz} offer higher compression ratios (e.g., 3.3$\times$ to 436$\times$ for SZ~\cite{7516069}) while preserving critical information within user-defined error bounds. Many lossy compressors utilize predictive techniques, such as curve-fitting~\cite{7516069} or spline interpolation~\cite{lakshminarasimhan2013isabela}, to exploit data correlations and smoothness, thereby reducing entropy and improving compression efficiency. However, spiky or irregular data, common in scientific simulations, disrupt these correlations, leading to higher prediction errors. These errors decrease the compression efficiency and increase storage requirements, motivating us to design more robust data compressors to handle data irregularities and keep improving compression efficiency.

Deep neural networks (DNNs), known as powerful predictor~\cite{hornik1989multilayer}, have achieved remarkable success in tasks like image classification, object detection, and super-resolution~\cite{rawat2017deep, liu2020deep, kong2022human, wang2020deep, torfi2021naturallanguageprocessingadvancements, anwar2020deep}, and researchers have explored their potential to enhance compression quality~\cite{64955684804448d0bf69755fb8f9dc08, d732dc4b721d438c820827a2abe06e8f, 10820597, li2024attention, gwlz}.
However, core challenges exist to hinder DNN application in scientific computing at ease.
\Circled{1} Static model training: Models are trained offline on fixed datasets, which poses significant limitations in the context of scientific data compression. Scientific data often originates from diverse domains across varying spatial and temporal scales~\cite{9378449}, leading to substantial distribution shifts between training datasets and real-world scenarios. Consequently, models pre-trained on static datasets struggle to generalize on previously unseen data, resulting in poor predictive performance when deployed~\cite{moreno2012unifying}.
\Circled{2} Attempted mitigation of data shift: To minimize the impact of data shift, researchers pre-trained large models on diverse datasets, aiming to bridge the distribution gap. However, this approach incurs high pre-training costs and increases user-side download overhead due to the sheer size of these models (e.g., HAT with 9 million parameters~\cite{d732dc4b721d438c820827a2abe06e8f} and Auto-Encoder with 1 million parameters~\cite{64955684804448d0bf69755fb8f9dc08}). Despite these efforts, data shift remains unresolved, as no model can comprehensively cover all possible distributions in scientific domains.
\Circled{3} Resource constraints: The large size of these models imposes significant computational demands on resource-constrained hardware during deployment, making their use for decompression inference challenging in practical scenarios.

This paper introduces \proj, a novel method to enhance the quality of error-regulating scientific lossy compression using online learning. %
Instead of relying on offline-trained large models toward generalization to reconstruct data, we integrate the process of finding data features into the scientific compression workflow in a finer-grain online manner, which {\it orthogonally} enhances a conventional compression method rather than substituting it.
More specifically, {\proj} is a quality enhancer that leverages the learned error residual features to improve the quality of the reconstructed data from the conventional method during decompression.

Effective enhancement of the conventional data reconstruction entails a reduction in the compression error, i.e., the difference between the reconstructed data and the original data.
Figure~\ref{fig:error-anatomy} (top) displays the error magnitude changes before and after the enhancement, with the latter smaller in error magnitude within the error bound. Also, the error residual between the conventionally reconstructed data and its enhancement is ``memorized'' in a DNN model.
We find that the reconstructed-enhanced error residual can be encoded using only thousands of parameters (e.g., 3,000).
Despite seemingly a very small number of parameters, it efficiently captures the fine details and complex patterns in the compression error overlooked by conventional compressors.
This method is much more effective than ``memorizing'' reconstructed-original error with full fidelity at an unknowingly high cost; it is also three orders of magnitude lower than previous DNN compression work in parameters (millions).

To do so, during compression, we online train a lightweight DNN model for each input data field,  which ``memorizes'' the features of the residual between the reconstructed and original data.
The method is at a per-block granularity, and the learned data feature is directly saved to the compressed file and is only applied to the input data.
The core innovation of \proj~is to \ul{\textit{handle data shift through fast and adaptive online learning of lightweight skipping DNN models}}. These models learn residuals between decompressed and original data, integrating cross-field learning with error management. Unlike usually practiced DNN-based compressors~\cite{64955684804448d0bf69755fb8f9dc08, d732dc4b721d438c820827a2abe06e8f}, which struggle with data shift, \proj~ dynamically adapts to changing data characteristics, making it highly effective for diverse scientific datasets. Designed as lightweight enhancers, the skipping DNN models minimize storage and computational overhead while improving decompression quality with minimal resource impact.

\begin{figure}[t]
  \includegraphics[width=\linewidth, trim={0 .45in 0 0}]{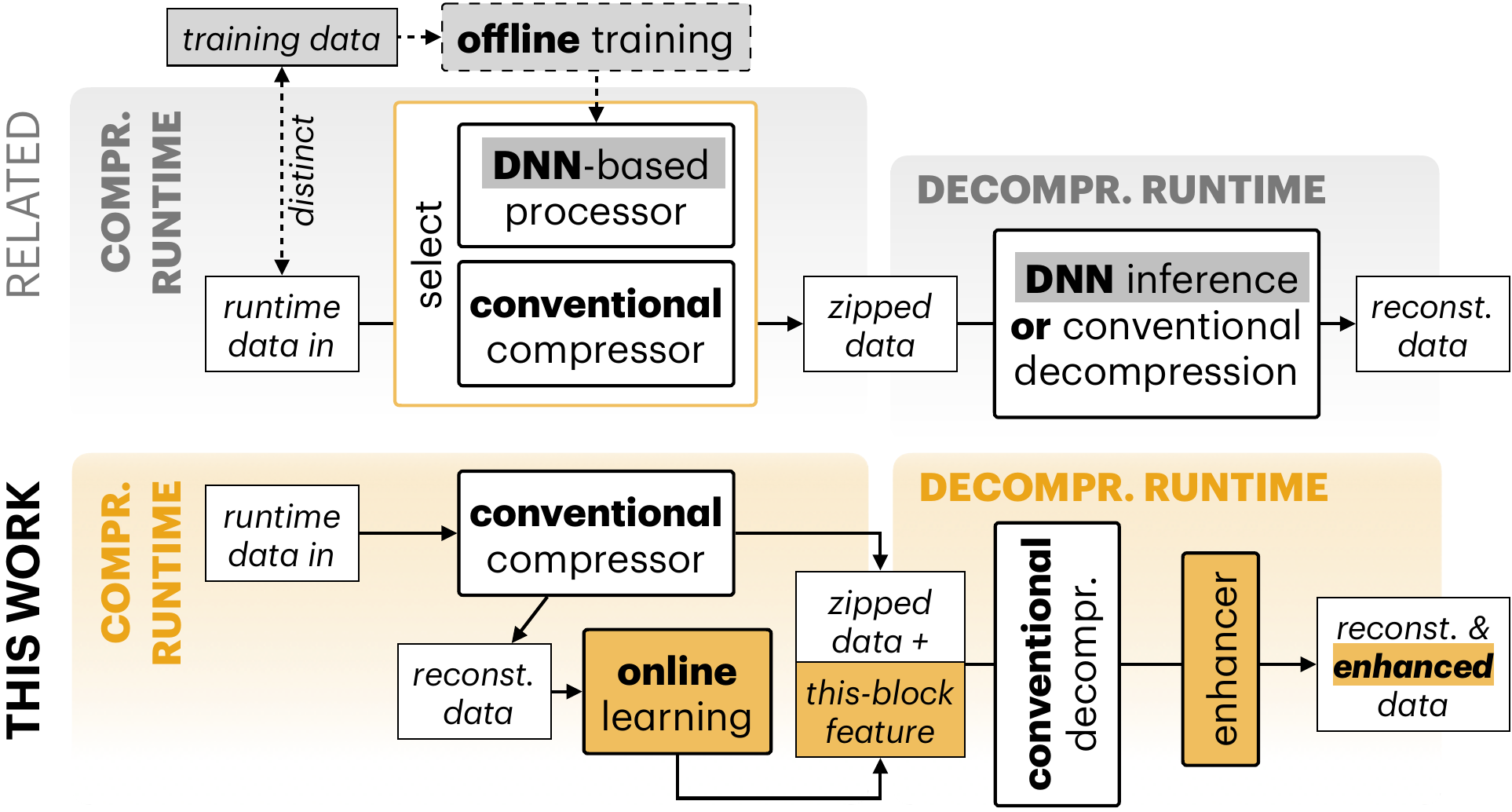}
  \caption{The {\proj} workflow ({\it bottom}) compared with related work consisting of a conventional compressor and alternative DNN-based processor ({\it top}).}
  \label{fig:overivew}
  \vspace{-\baselineskip}
\end{figure}

The key contributions are summarized as follows:

\begin{itemize}[leftmargin=1.3em]
    \item A novel method, \proj, to enhance the quality of scientific lossy compression with online neural learning. \proj~ dynamically trains lightweight skipping DNN models during compression, enabling efficient adaptation to data shift. \textcolor{black}{We design the method to be theoretically adaptive to any conventional lossy compressors for quality enhancement.}
    \item \textcolor{black}{Compression error mitigation: \proj~efficiently mitigates compression errors introduced by scientific compressors. By employing lightweight skipping DNN models, \proj~introduces little space and time overhead to the original compression process while significantly enhancing the quality of reconstructed data.}
    \item \textcolor{black}{Error control:  our framework introduces two regulation modes, allowing users to choose from strict error control with respect to the user-input error bound for the compressor and the relaxed error control that achieves overall better quality with worst-case capped at $2\times$ the user-input error bound.}
    \item Cross-field learning: We leverage cross-field correlations derived from governing equations in scientific simulations to improve model predictions. By incorporating data from multiple fields, the framework captures cross-field dependencies often overlooked by traditional lossy compressors, further improving reconstruction quality. The cross-field learning can be adopted based on data features in various scientific applications.
    \item Extensive evaluation: \proj~ has been evaluated on multiple datasets and compared against state-of-the-art lossy compressors like SZ3~\cite{liang2022sz3} and ZFP~\cite{lindstrom2014fixed}, demonstrating significant performance gains.
\end{itemize}

\section{Background and Motivation}

\subsection{Lossy Compression}
\label{section_lossy-compression}

Conventional error-bounded lossy compressors fall into several types, including prediction-based, transform-based, and HOSVD-based approaches \cite{di2024surveyerrorboundedlossycompression}.
Prediction-based compressors \cite{lindstrom2014fixed, 7516069, 10046076, 10.1145/3639259, 9458791, 7967203, liang2021mgard+, 10.1145/3295500.3356193, tao2019optimizing, 10.1145/3369583.3392688, 8622520} use predictors to estimate pointwise data based on neighboring values, followed by quantization to maintain user-defined error bounds. Data prediction is the most critical stage in these compressors, as higher accuracy reduces the burden on subsequent steps~\cite{di2024surveyerrorboundedlossycompression}. Existing methods employ predictors such as curve fitting~\cite{7516069}, spline interpolation~\cite{lakshminarasimhan2013isabela}, multidimensional prediction~\cite{7967203}, and higher-order predictors~\cite{10.1145/3369583.3392688}.
Transform-based approaches \cite{lindstrom2014fixed, li2023lossy} operate by converting the data into a coefficient space that is more amenable to compression due to its sparsity. However, maintaining control over the error within predefined bounds in this domain is often challenging.
HOSVD-based techniques \cite{ballester2019tthresh, ballester2015analysis} leverage Higher-Order Singular Value Decomposition (HOSVD) to decompose data into matrices and a compact core tensor, ensuring that the L2 norm error is preserved. While these methods excel in achieving high compression ratios by capturing global correlations in the data, they are significantly limited by their slow computational performance.

To push the limit of conventional lossy compression, recent efforts ~\cite{64955684804448d0bf69755fb8f9dc08, d732dc4b721d438c820827a2abe06e8f, 10820597, li2024attention} attempt to leverage deep neural networks (DNNs) to predict the reconstructed data adaptively, as a learnable predictor trained offline and integrated into existing prediction-based lossy compressors.

\subsection{Deep Neural Networks}
\label{section_dnn}

Deep neural networks (DNNs) are powerful machine learning models composed of hierarchical neural layers that extract increasingly complex features from input data. DNNs have achieved remarkable success in a wide range of real-world applications \cite{9025499, Fan_2022, torbunov2022uvcgan, 10.1145/3240508.3240625}. However, a persistent challenge in deploying DNNs lies in their generalization problem -- the gap between their performance on the training data and their prediction performance on unseen data \cite{Lust2020ASO, ROHLFS2025128701}. This issue is exacerbated by data shift, where the distribution of the training data diverges from that of the deployment data~\cite{moreno2012unifying}. DNNs are fundamentally designed under the assumption that training and deployment data are drawn from the same independent and identically distributed (i.i.d.) distribution. When this assumption breaks, as is common in real-world scenarios, their learned representations often fail to adapt, leading to unreliable and suboptimal results. Addressing these challenges requires approaches that not only mitigate the impact of distribution shifts but also dynamically adapt to evolving data environments to maintain consistent performance.

\begin{figure*}[t!]
  \centering
  \begin{minipage}[t]{0.49\textwidth}
    \includegraphics[width=\linewidth]{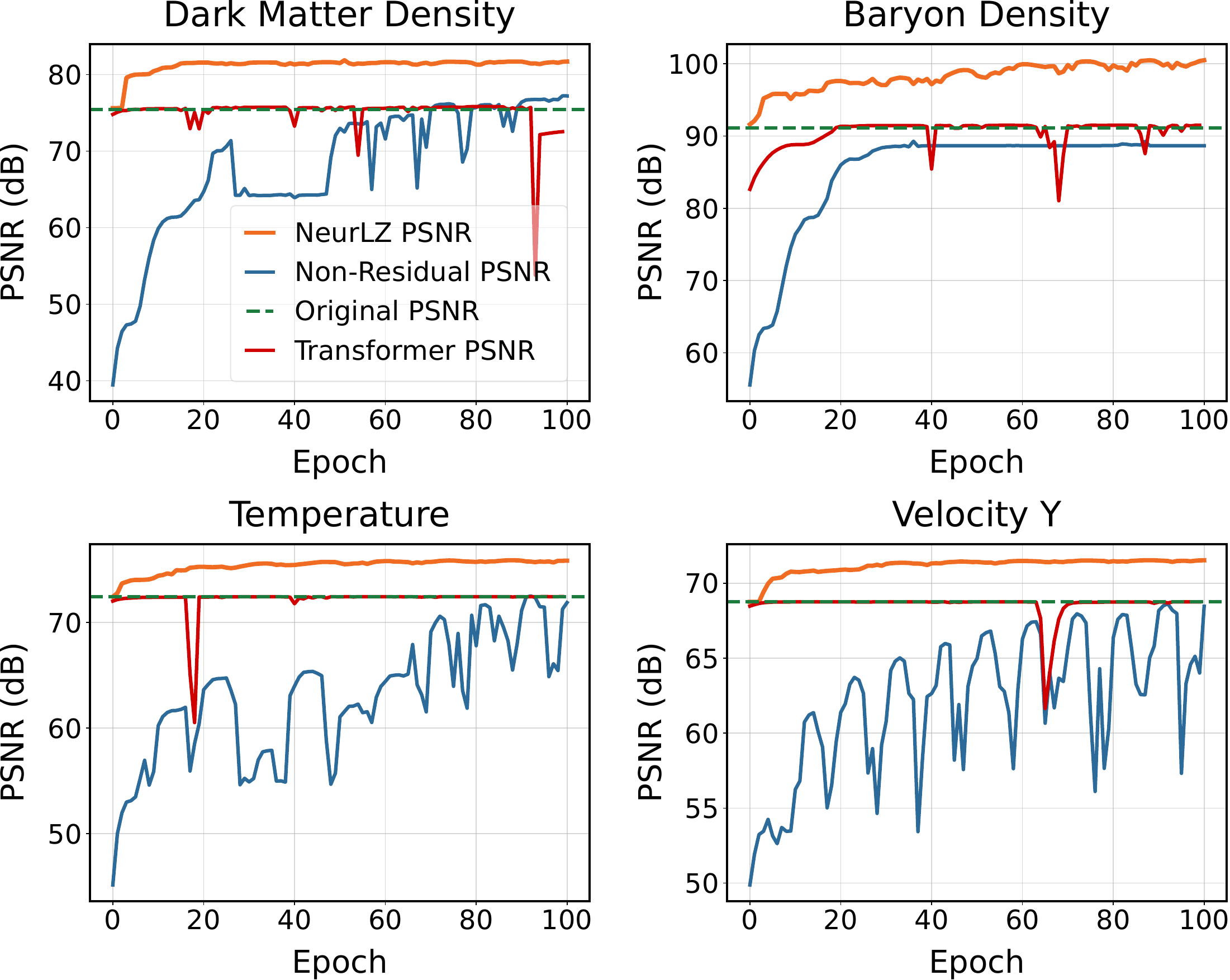}
  \end{minipage}~%
  \begin{minipage}[t]{0.49\textwidth}
    \includegraphics[width=\linewidth]{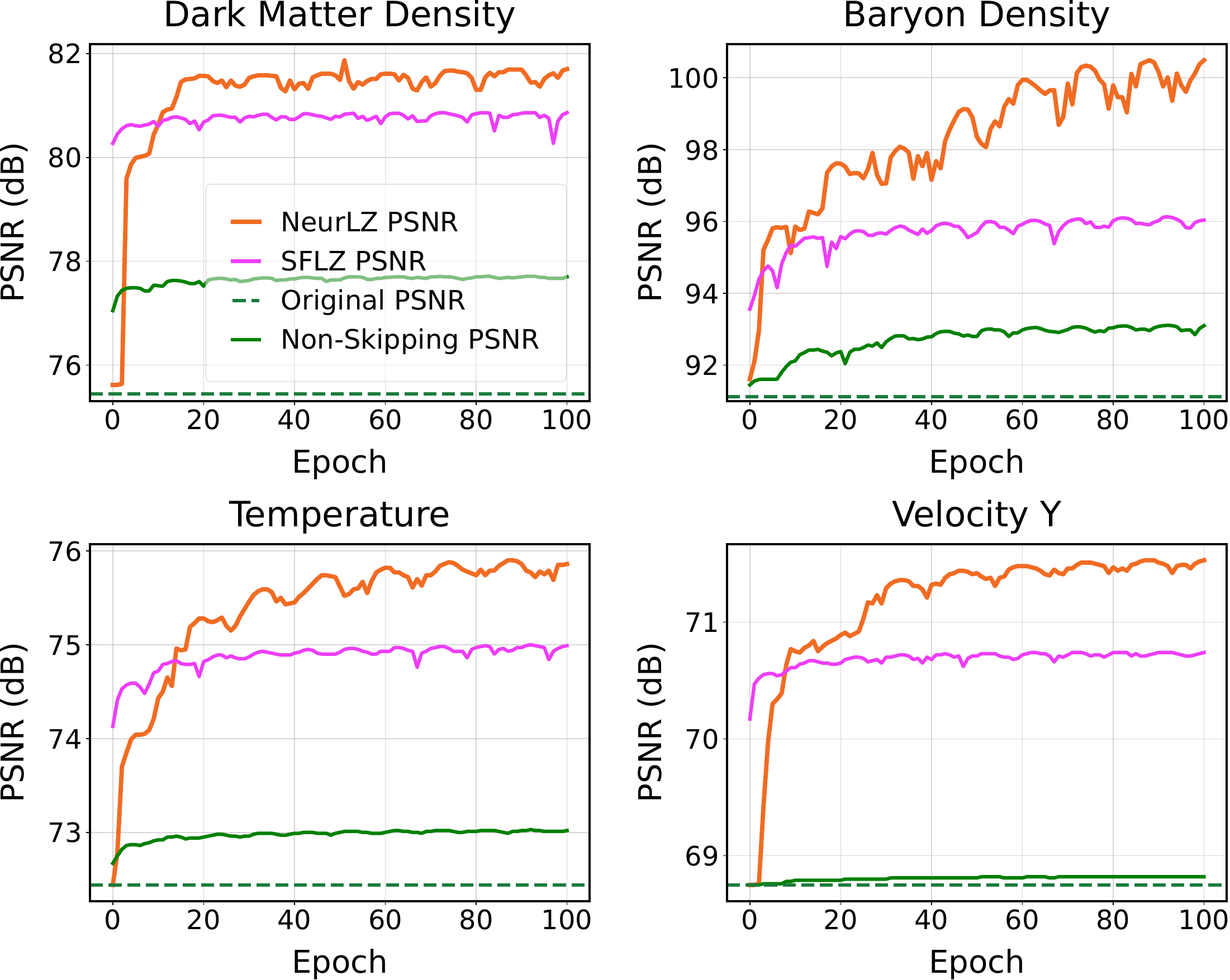}
  \end{minipage}
  \vspace{-0.5\baselineskip}
  \caption{\proj~achieves higher PSNR across all Nyx fields, outperforming both the non-residual and transformer models (left), as well as the non-skipping and single-field learning (\sproj) models (right).}
  \Description{A figure comparing PSNR of \proj, non-residual models, transformer case, and \sproj models across all Nyx fields.}
  \label{fig:effect}
  \vspace{-0.5\baselineskip}
\end{figure*}

\subsection{Motivation}
\label{Section_motivation}
Our ultimate goal is to fully unlock the power of DNNs to push the limit of scientific lossy compression quality. In this subsection, we comprehensively rethink the relationship between the utilization of DNNs and the principles of lossy compression. Specifically, we investigate the optimal design philosophy to integrate DNN models by analyzing and answering the following two critical questions.

\begin{formal}
  \textit{Question \Circled{1}: What is the best learning regime for lossy compression? (Offline Learning vs. Online Learning.) }
\end{formal}
\vspace{-.3\baselineskip}

Existing offline-learning methods~\cite{64955684804448d0bf69755fb8f9dc08, d732dc4b721d438c820827a2abe06e8f, 10820597, li2024attention} feature a large DNN model pre-trained on a pre-collected static dataset and then deployed for inference.
However, this approach heavily assumes that training and deployment data share the same underlying distribution. In practice, this assumption rarely holds, as scientific domains frequently involve complex systems with significant distribution variance across fields like cosmology simulations (e.g., Nyx~\cite{almgren2013nyx}), large-scale turbulence (e.g., Miranda~\cite{miranda_simulation}), and weather simulations (e.g., Hurricane~\cite{hurricane-data}). Each domain has distinct data distributions, and even within a single domain, the data can vary substantially across spatial or temporal contexts~\cite{9378449}. For instance, Nyx exhibits temporal variations corresponding to different stages of the universe's evolution indicated by redshift \cite{almgren2013nyx}. These temporal shifts result in fundamentally different data distributions as physical phenomena evolve and dominate at distinct epochs. Consequently,
offline learning
can be ill-equipped to adapt to such temporal dynamics, further amplifying the impact of data shift on model performance.

Online learning~\cite{hoi2021online, CesaBianchi2006PredictionLA, shalev2012online, JMLR:v15:hoi14a} addresses this issue by allowing models to dynamically update their parameters or replace themselves as new data becomes available. This continuous updating approach ensures the model stays aligned with evolving data distributions, maintaining robust performance over time. Research in real-time psychographic imaging~\cite{babu2023deep} and autonomous X-ray reflectometry experiments~\cite{pithan2023closingloopautonomousexperiments} highlights the potential of online learning to address data shift problems across scientific domains.

Recognizing these advantages, \proj~adopts online learning as the core regime to tackle the challenges posed by data shift in scientific lossy compression. By leveraging the adaptability of online learning, \proj~ensures consistent and high-quality compression, making it well-suited for dynamic and diverse scientific scenarios. This approach represents a significant step toward overcoming the limitations of traditional offline learning and addressing the complexities of real-world datasets.

\begin{formal}
  \textit{Question \Circled{2}: What is the appropriate learning granularity for scientific data compression? (Generalization vs. Customization.)}
\end{formal}
\vspace{-.3\baselineskip}

Generalization is a potential online learning strategy that trains a huge DNN model in the compression process, handling the entire target data. While this approach can "memorize" a wide range of features, it incurs significant computational and inference costs, requiring substantial resources for training, deployment, and user-side inference. Moreover, downloading and maintaining such a model adds significant I/O and storage overhead, leading to impracticality and unscalability in HPC applications.

On the contrary, customization employs multiple lightweight DNN models that deal with individual small data blocks. These models learn per-block data mappings, capturing the unique characteristics of each block. Compared to a general huge model, customized lightweight models benefit from lower computational costs, reduced inference resource requirements, and minimal download overhead for users.
\proj~adopts this customization online learning strategy, ensuring high-quality compression performance and resource efficiency by updating multiple block-specific lightweight DNN models online in the lossy compression process.

\section{Methodology: \texttt{NeurLZ}}
\label{overview}

\subsection{Method Overview}
\label{section-design-overview}

As highlighted in Sec.~\ref{Section_motivation}, \proj~distinguishes itself from prior works by employing online neural learning and enhancing the quality of reconstructed data using learned DNN enhancer models in the post-processing stage, as shown in Figure~\ref{fig:overivew}. \proj~dynamically learns multiple lightweight DNN models for each runtime data block in a customized way.
Specifically,
when a runtime data block arrives, a lightweight DNN model is initialized for online learning in the compression process.
\proj~leverages the decompressed data as input and trains the lightweight model to predict the residual error (i.e., the difference between the original and reconstructed data). This process trains DNN models to minimize the discrepancy between decompressed and original data, effectively refining the decompressed data to recover the original values.

Once the DNN encodes the error residuals (Figure~\ref{fig:error-anatomy}), the model weights (negligible size compared to compressed data) and outlier coordinates (points exceeding 1$\times$ error bound, detailed in Sec.~\ref{section_error-bound}), are packaged as the specific
feature for this runtime data block along with the zipped data. This packaged format is then sent to the user.
On the user side, the reconstruction process begins with conventional decompression to obtain reconstructed data. The saved lightweight DNN models and the outlier coordinates are used to enhance data quality in the post-decompression stage. Specifically, the DNN models predict the residual error, which is then added to the reconstructed data to generate enhanced data block by block. For data points in the outlier coordinates, the original reconstructed values are retained, as will be detailed in Sec.~\ref{section_error-bound}. This ensures the enhanced data achieves high fidelity while adhering to the user-defined error bound.

\subsection{Lightweight Online Neural Learning}
\label{section_skippingDNNs}

As discussed in Sec. \ref{Section_motivation}, our \proj~contains multiple lightweight DNN models, which can be considered a memorizer to encode residual error information. Each memorizer is learned for a specific runtime data block, capturing its unique characteristics. The memorizer must balance
feature-extracting capability and size:
a weak model provides low-quality enhancement, while a large model suffers from significant overhead by outweighing the compressed data. We will introduce the detailed learning design of our \proj.

\subsubsection{Learning Residual Error Information. }
A key challenge in \proj's learning process is enhancing decompressed data to recover the original values, especially given the large value ranges in scientific simulations, such as the $4.78\times 10^6$ range in Nyx's Temperature field~\cite{9378449}, which causes training instabilities.
Instead of directly predicting the original data, \proj~ estimates the residual error, $R=X-X'$, inspired by residual learning techniques~\cite{zhang2017beyond, 8237748, 10.1007/978-3-030-01264-9_19}. By treating data slices as single-channel images, the DNN model focuses on smaller residuals, stabilizing training and improving prediction accuracy. This approach effectively enhances decompressed data to maximally recover the original data, making the DNN a strong feature-capturing memorizer for each runtime data block.

The four left subfigures in Figure~\ref{fig:effect} compare learning residual errors to learning original values for the Nyx dataset compressed with SZ3 under a relative error bound of 1E-3. PSNR (Peak Signal-to-Noise Ratio) in decibels (dB) is used to
measure compression-introduced errors; higher PSNR indicates better reconstruction quality. \proj~(orange solid line) consistently achieves higher PSNR and improves reconstruction quality with learning residual error. In contrast, directly learning original values (green solid line) shows lower and less stable performance. These results demonstrate the effectiveness of the proposed residual-error learning strategy in improving reconstruction quality.

Upon completing the learning process during \proj-compression, 
the trained DNN model predicts the residual error map $\hat{R}$ directly from the decompressed data $X'$ through learned correlations. The predicted error is then added back to $X'$, further enhancing the overall reconstruction quality as $\hat{X}=X'+\hat{R}$.

\subsubsection{Learning Multi-Scale Patterns with Skipping Neural Networks. }
The skip connection structure empowers DNNs to represent intricate patterns spanning various scales \cite{he2015deepresiduallearningimage, 10.1007/978-3-319-24574-4_28, 9025499, Fan_2022, torbunov2022uvcgan, 10.1145/3240508.3240625}. Accordingly, we design lightweight skipping DNNs, as shown in Figure~\ref{fig:model}. The input slices from decompressed blocks undergo down-sampling and up-sampling through convolution and de-convolution to extract multi-scale features. Skip connections concatenate low-level features with deeper layers, enriching the representation. The fused features produce a single-channel output as the predicted residual map, which, when added to the decompressed slice, yields the enhanced slice via \proj.

\begin{figure}[t!]
  \centering
  \includegraphics[width=\linewidth]{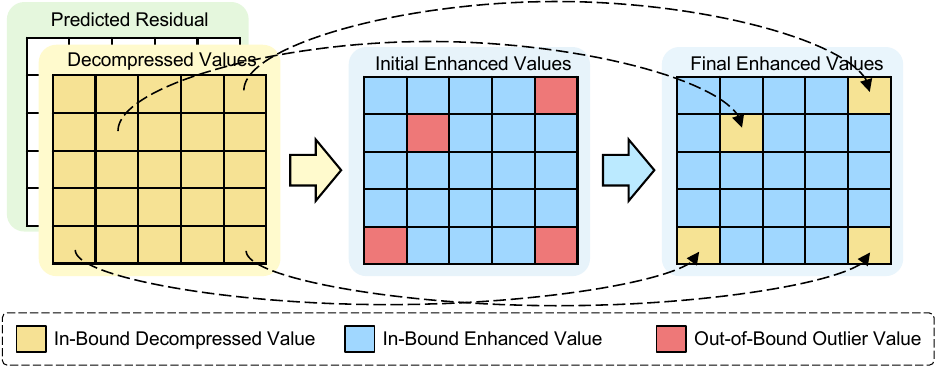}
  \vspace{-7mm}
  \caption{Outlier management: decompressed (reconstructed) values are further enhanced with predicted residuals, and outliers are replaced with in-bound decompressed values to reliably satisfy the error bound.}
  \Description{Diagram showing enhanced decompressed values with residual predictions and outliers replaced to meet error bounds.}
  \label{fig:outlier}
  \vspace{-1.5\baselineskip}
\end{figure}

Leveraging the skip connection structure's ability to capture multi-scale information, our skipping DNNs effectively model complex patterns while preserving high-fidelity details. As shown in Figure~\ref{fig:effect}-right, the non-skipping model (green solid line) achieves lower PSNR compared to our \proj~model (orange solid line) with skip connections. Beyond improving reconstruction quality, the integration of convolution and de-convolution operations ensures model compactness; for instance, a 10-layer network requires only 3,000 parameters. This lightweight design makes the memorizer particularly suitable for transmission, especially when compared to billion-parameter Transformer models%
~\cite{d732dc4b721d438c820827a2abe06e8f, 10.1145/3485447.3511987}.

Furthermore, as shown on the left side of Figure~\ref{fig:effect}, we compare our skipping model with the Transformer-based HAT model~\cite{chen2023hat}, designed for super-resolution tasks. Despite its 5 million parameters---far exceeding our 3,000-parameter skipping model -- HAT struggles to improve reconstruction quality, even excluding its longer training time. This is due to Transformers being more data-hungry, making them more challenging to learn on small datasets~\cite{gani2022small,has2022escap,liu2021efftrans,9944625,xu2021opttrans}. In our case, the original data block size is $512^3$, equivalent to 512 images whose sizes are 512-by-512---a very small dataset by computer vision standards. In contrast, our memorizer design, which incorporates skip connections, achieves better results with significantly smaller data requirements. Its compact and effective nature perfectly aligns with \proj's goals for enhancing scientific data compression quality.

\begin{figure}[t!]
  \centering
  \includegraphics[width=0.95\linewidth, trim={0 .1in 0 0}]{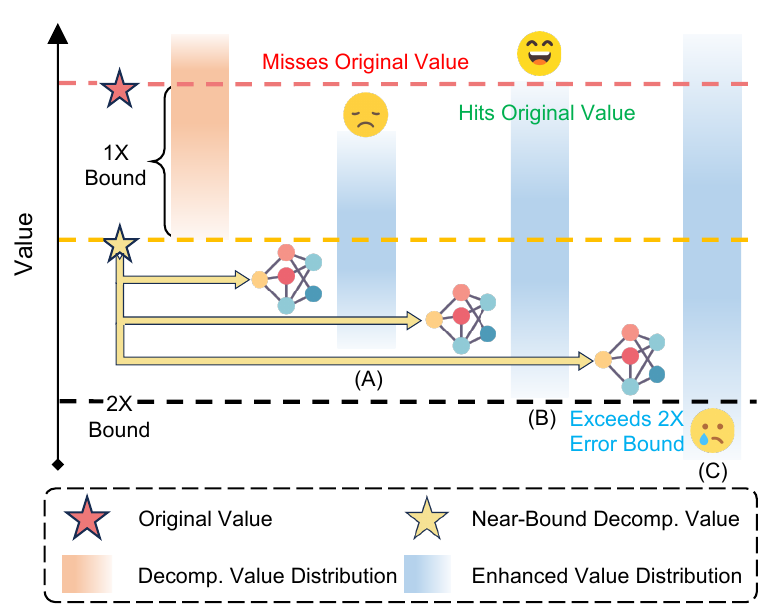}
  \vspace{-1mm}
  \caption{Regulated DNN output: The red band shows decompressed values, with the worst-case value (yellow star) processed by three DNNs—tight (A), balanced (B), and loose (C) regulation. The blue bands show that Case A misses the original, Case B reaches it accurately, and Case C exceeds the 2$\times$ bound.}
  \Description{Visualization of DNN regulation.}
  \label{fig:2xbound}
  \vspace{-\baselineskip}
\end{figure}

\subsection{Flexible and Strict Error Bounding}
\label{section_error-bound}
\subsubsection{Strict Error Control. }

The inherent uncontrollability of DNN output causes some data points, even after enhancement with predicted residuals, to exceed the error bound. These points, referred to as outliers, require special handling to guarantee the error-bound.
For the strict error-bounded reconstruction enhancement, we store the coordinates of outlier points, defined as those whose enhanced values exceed the acceptable error threshold (Figure~\ref{fig:outlier}-left). During reconstruction, these coordinates are used to identify and replace out-of-bound enhanced values with their corresponding decompressed values, ensuring all data points remain within the predefined error limits (Figure~\ref{fig:outlier}-right).
This strategy leverages DNNs' ability to capture complex patterns while addressing output variability, preserving accuracy and reliability. However, storing outlier coordinates adds additional storage overhead.
We denote the size of the $i$-th dimension as dim$_i$ and the average number of bits required to store the $N$-D coordinates of a single point as $\overline{B}$. With $\overline{B} = \textstyle {\sum_{i=1}^{N} \log_2(\operatorname{dim}_i)}$,
the coordinate overhead of storing outliers in bits is obtained by the number of outliers multiplied by $\overline{B}$.
For the datasets discussed in this paper, the $\overline{B}$ values are as follows: Nyx has $\overline{B}$ values of 27.0 bits for size$=512^3$, 30.0 bits for size $=1024^3$, and 33.0 bits for size$=2048^3$; Miranda has a $\overline{B}$ of 25.2 bits; Hurricane a $\overline{B}$ of 24.6 bits.

\subsubsection{Regulating Neural Output. }
In some cases, storing the coordinates of outliers incurs excessive overhead, as shown in Sec.~\ref{section_discussion_no_outlier}. We propose a regulating technique for application users who need an extremely high compression ratio to achieve a 2$\times$ error bound per datum without storing outlier coordinates. As depicted in Figure~\ref{fig:2xbound}, consider a single datum: the original value (red star \textcolor{red!80!black}{$\star$}) is compressed and decompressed within a 1$\times$ bound by lossy compressors, with the decompressed value distribution shown as the red band. The worst-case decompressed value (yellow star \textcolor{yellow!80!black}{$\star$}) lies at the bound, 1$\times$ from the original. This value is input into the DNN, producing an enhanced value within the range shown as the blue band. While the DNN's output cannot be precisely controlled, its range can still be effectively regulated.

\begin{figure}[t!]
  \centering
  {\includegraphics[width=\linewidth, trim={0 -.4in 0 0}]{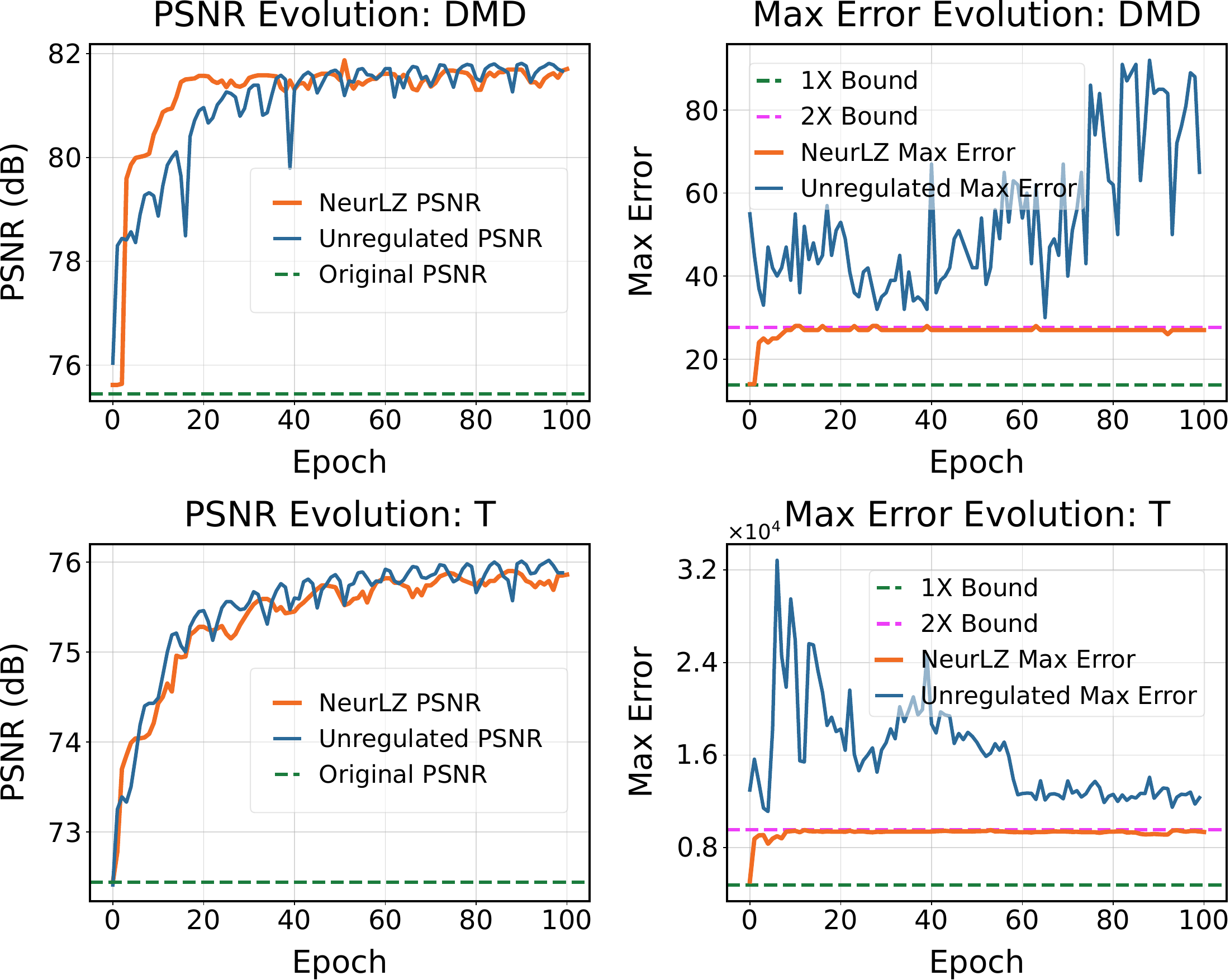}}
  \vspace{-2\baselineskip}
  \caption{Training evolution of PSNR and max absolute error for two fields in the Nyx---\proj~ (regulated) vs. unregulated cases. \proj~ maintains stability within the 2$\times$ bound, while the unregulated case exceeds it.}
  \Description{Figure showing PSNR and max absolute error evolution for two Nyx fields, comparing \proj-regulated and unregulated cases.}
  \label{fig:2xbound-evolution}
  \vspace{-1\baselineskip}
\end{figure}

Figure~\ref{fig:2xbound} shows three scenarios with varying degrees of DNN output regulation. In Case A, tight regulation keeps the enhanced value within the 2$\times$ bound, but the maximum falls short of the original, potentially limiting performance. Case B achieves balanced regulation, with the minimum meeting the 2$\times$ bound and the maximum reaching the original, allowing the DNN to recover the original value through learning. In Case C, loose regulation allows the maximum to exceed the original, but the minimum also surpasses the 2$\times$ bound, risking degraded reconstruction quality.

In \proj, balanced regulation normalizes residuals with the error bound and uses a Sigmoid layer in the skipping DNN to constrain outputs to [0, 1], equivalent to a 2$\times$ bound. Figure~\ref{fig:2xbound-evolution} shows the training evolution of PSNR and maximum absolute error for two Nyx fields, comparing regulated \proj~and the unregulated case. Both enhanced SZ3 with a 1E-3 bound and achieved similar PSNR, but regulated \proj~ensures stable training. It consistently keeps errors within the 2$\times$ bound. By default, outlier coordinates are saved to enforce strict bounds, but omitting them improves compression ratios with minimal impact (Sec.~\ref{section_discussion_no_outlier}).

\subsection{Cross-Field Learning}
\label{section_group-wise}

Scientific applications model multiple physical fields in a spatiotemporal domain, such as temperature, velocity, and density in the Nyx simulation, governed by complex equations like \emph{Dark Matter Evolution} and \emph{Self-gravity Equations}~\cite{almgren2013nyx, lukic2015lyman, onorbe2019inhomogeneous}.
However, existing lossy compressors~\cite{7516069, lindstrom2017fpzip, 7967203, liang2022sz3, 10.1145/3369583.3392688, d732dc4b721d438c820827a2abe06e8f} are typically limited to take single-field inputs, thereby overlooking inherent physical connections among multiple fields and timesteps.

\begin{figure}[t!]
  \centering
  \includegraphics[width=\linewidth]{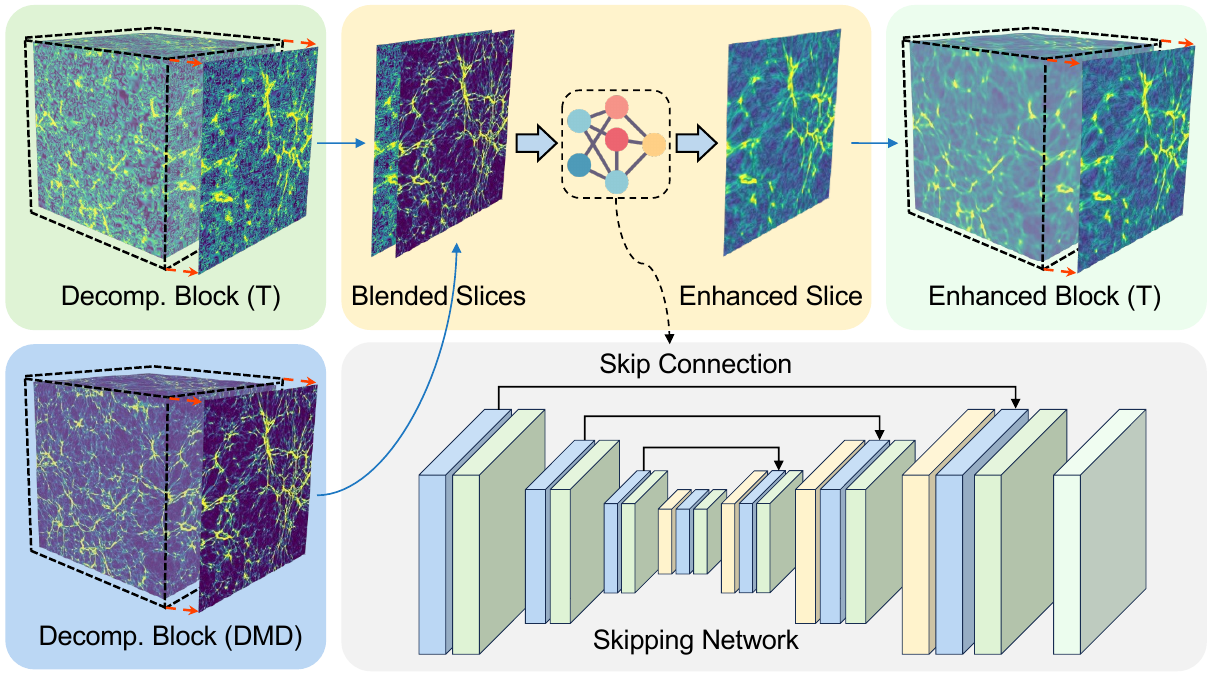}
  \vspace{-1\baselineskip}
  \caption{Skipping DNN model and cross-field learning. `Decomp.' denotes the decompressed data,  while `T' and `DMD' denote two fields where the former is enhanced.}
  \Description{Diagram of the learnable enhancer model. It includes decompressed data, two fields (T and DMD), and a process to enhance the first field.}
  \label{fig:model}
  \vspace{-1.5\baselineskip}
\end{figure}

\begin{figure}[t]
  \centering
  \includegraphics[width=\linewidth, trim={0 .2in 0 0}]{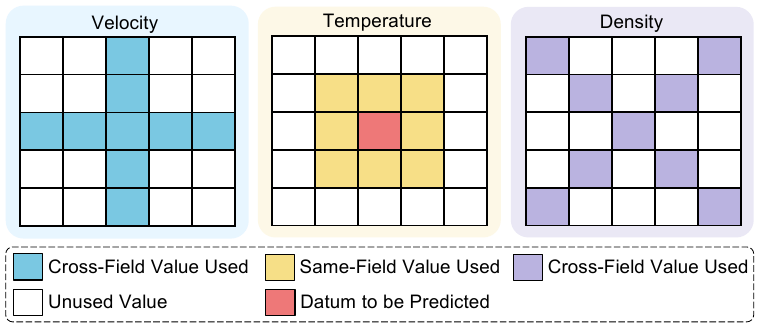}
  \caption{Cross-field learning: A datum’s prediction leverages values from its own field as well as other related fields through learnable patterns.}
  \Description{Illustration of cross-field learning where a datum’s prediction combines values from its own field and other related fields in a learnable manner.}
  \label{fig:cross-field-pattern}
    \vspace{-0.5\baselineskip}
\end{figure}

Using skipping DNNs to model complex inter-field patterns, we implement cross-field learning, as shown in Figure~\ref{fig:cross-field-pattern}. For instance, predicting a single pixel (red cube in the Temperature field) leverages local values and information from other fields. These patterns are learned during training, enabling predictions informed by correlations across fields. By capturing such relationships, the skipping model achieves a more accurate prediction of the data source, thereby enhancing compression ratios.

Figure~\ref{fig:model} illustrates the training process of a skipping model for cross-field learning. Decompressed slices from Temperature (T) and Dark Matter Density (DMD) serve as a 2-channel input to predict the residual map for Temperature. After fusing features across channels, the model outputs the residual map, which is added to the decompressed slice to produce an enhanced slice via \proj. The right side in Figure~\ref{fig:effect} highlights the advantages of cross-field learning. While single-field learning (purple solid line) provides some improvement over the original PSNR (green dashed line), it consistently underperforms compared to cross-field learning (orange solid line). This underscores the superior ability of cross-field learning to capture complex inter-field dependencies and enhance model performance.

\section{Evaluation}

\begin{figure*}[t!]
  \centering
  \begin{minipage}[t]{0.49\textwidth}
    \includegraphics[width=\linewidth]{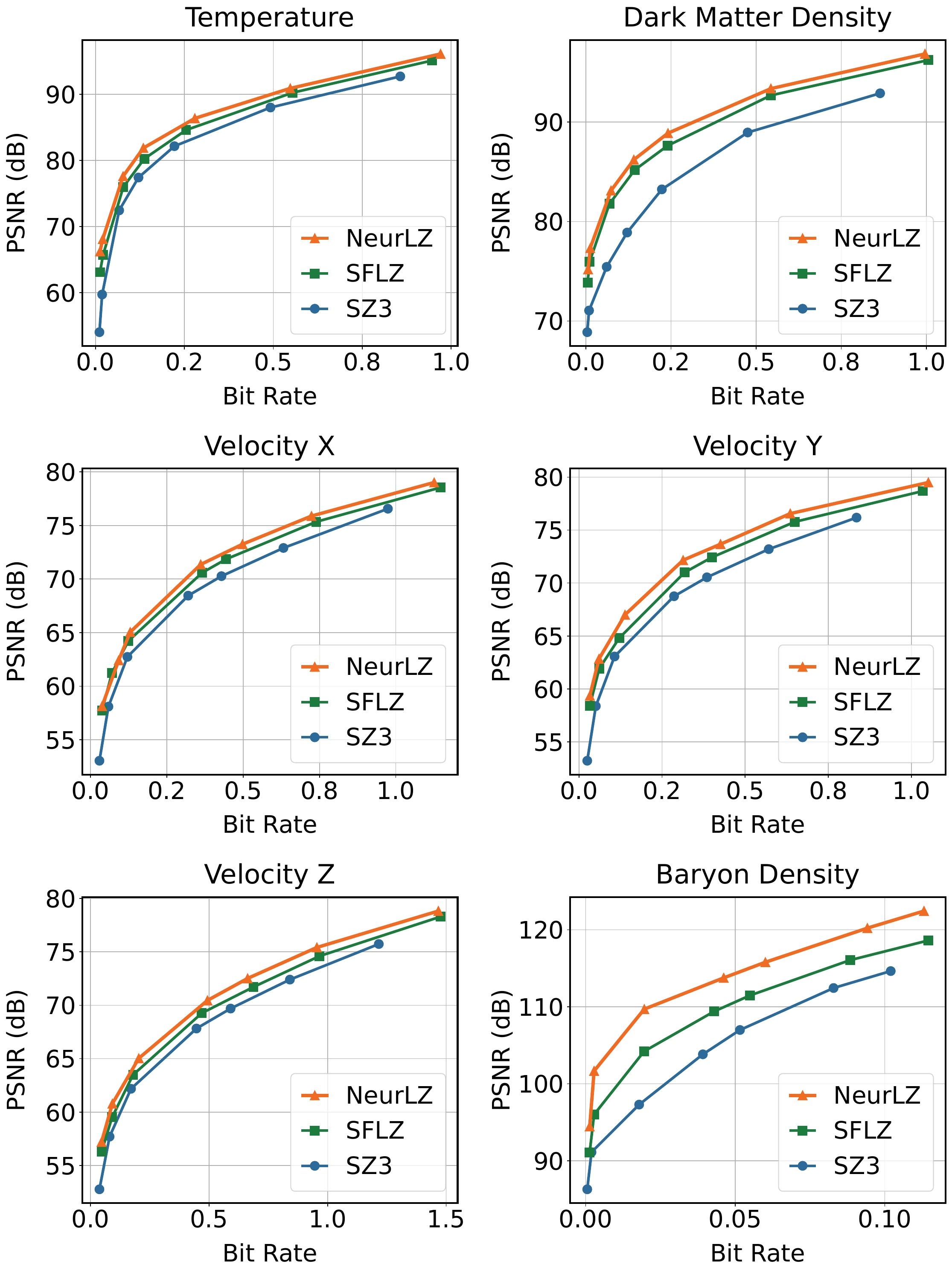}
  \end{minipage}
  \hfill
  \begin{minipage}[t]{0.49\textwidth}
    \includegraphics[width=\linewidth]{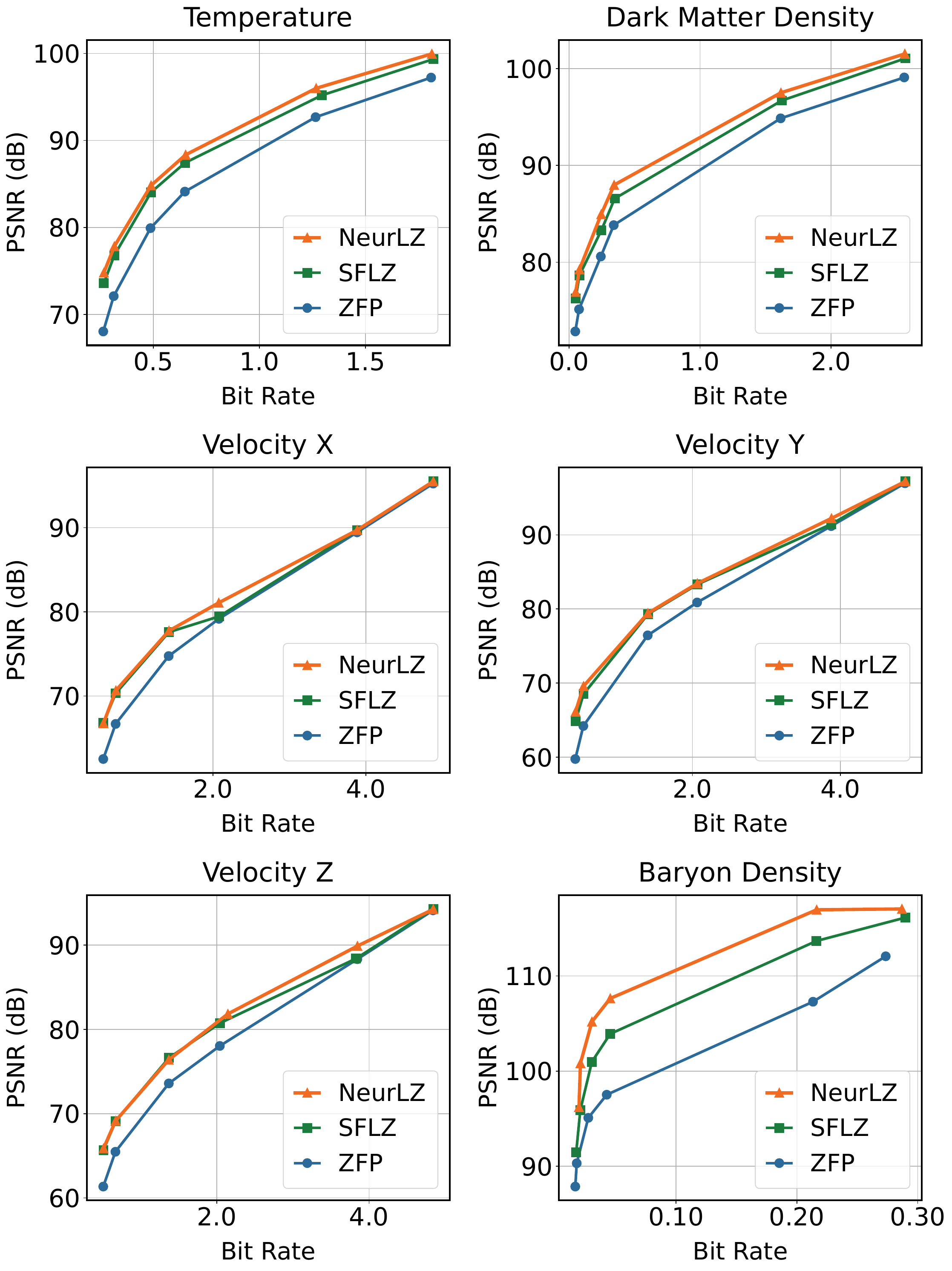}
  \end{minipage}
    \vspace{-0.5\baselineskip}
  \caption{Bit rate comparison for the Nyx dataset: (Left) SZ3-based methods—original SZ3, \sproj~ (single-field learning), and \proj~ (cross-field learning). (Right) ZFP-based methods—original ZFP, \sproj, and \proj.}
  \Description{Figure comparing bit rates for the Nyx dataset.}
  \label{fig:bitrate}
\vspace{-0.5\baselineskip}
\end{figure*}

In this section, we present the evaluation results of the proposed {\proj} on typical scientific applications compared to existing lossy compressors for scientific data.

\subsection{Experimental Setup}
\label{section_experiment-setup}

\FakeParagraph{Testbed} Experiments were conducted on the servers with eight NVIDIA RTX 6000 Ada GPUs, two 24-core AMD EPYC Genoa 9254 CPUs, and 1.5 TB of DRAM each.

\FakeParagraph{Datasets} Table \ref{tab:dataset} summarizes the three datasets used in our experiments: Nyx~\cite{almgren2013nyx}, Miranda~\cite{miranda_simulation}, and Hurricane~\cite{hurricane-data}. Sample blocks for Miranda and Hurricane were obtained via SDRBench~\cite{9378449}. Sample dataset of Nyx was sourced from SDRBench and NERSC open-data portal~\cite{nersc-data} for scalability-related experiments.

\FakeParagraph{Learning configurations} Each skipping DNN model comprises four down- and up-sampling operations with skip connections, totaling around 3,000 parameters. The training was conducted over 100 epochs with a batch size of 10, an initial learning rate of 1E-2, and a cosine annealing schedule. Once trained, the model weights are embedded into the final compressed data format in FP32 or FP64 precision, aligned with the dataset's precision.

\FakeParagraph{Lossy compressor} \proj~works seamlessly with various lossy compressors. We select two popular ones: the prediction-based SZ3~\cite{liang2022sz3} and the transform-based ZFP~\cite{lindstrom2014fixed}. In prediction-based compressors like SZ3, superior predictors effectively capture correlations. \proj~enhances this process by capturing previously overlooked correlations. For ZFP which uses orthogonal transforms to decorrelate data~\cite{di2024surveyerrorboundedlossycompression}, \proj~evaluates its performance with transform-based compressors.

\FakeParagraph{Performance evaluation criteria} Using compressor-specified error bounds, \proj~enhances decompressed values. In SZ3, we use a value-range-based relative error bound (denoted as $e$), which is equivalent to the absolute error bound $\epsilon$ used in ZFP. The relationship between them is given by \(e = \epsilon \cdot value\_range\), where $value\_range$ represents the specific data block's value range being compressed. For simplicity, all error bounds are referred to as $e$.  Experiments are defined by the compressor type, error bound, dataset, and field, and \proj~is evaluated using the criteria described below:

\begin{itemize}[leftmargin=1.3em]
  \item Error validation: verifying that \proj's error is strictly bounded by the user-defined error threshold.
  \item Data visualization: assessing the visual quality of enhanced data relative to the original decompressed data.
  \item Bit rate vs. PSNR: comparing PSNR across bit rates.
  \item Bit rate reduction: evaluating \proj's reduction relative to SZ3 and ZFP at equal PSNR.
  \item Scalability testing: testing \proj~on larger datasets.
\end{itemize}

\begin{table}[t!]
  \caption{Experiment-used datasets~\cite{9378449}.}
  \vspace{-.5\baselineskip}
  \label{tab:dataset}\footnotesize
  \renewcommand{\arraystretch}{1.2}
  \begin{tabular}{|c|c|c|c|}
    \hline
    \textbf{Dataset}                     & \textbf{Domain}  & \textbf{Per Block Size}   & \textbf{Data Type} \\
    \hline
    Nyx~\cite{almgren2013nyx,nersc-data} & Cosmology        & $512^3$/$1024^3$/$2048^3$ & FP32               \\
    \hline
    Miranda~\cite{miranda_simulation}    & Large Turbulence & (256, 384, 384)           & FP64               \\
    \hline
    Hurricane~\cite{hurricane-data}      & Weather          & (100, 500, 500)           & FP32               \\
    \hline
  \end{tabular}%
\end{table}

\subsection{Experimental Results}
\label{section_experimental-results}

\begin{figure}[t]
  \centering
  \includegraphics[width=\linewidth]{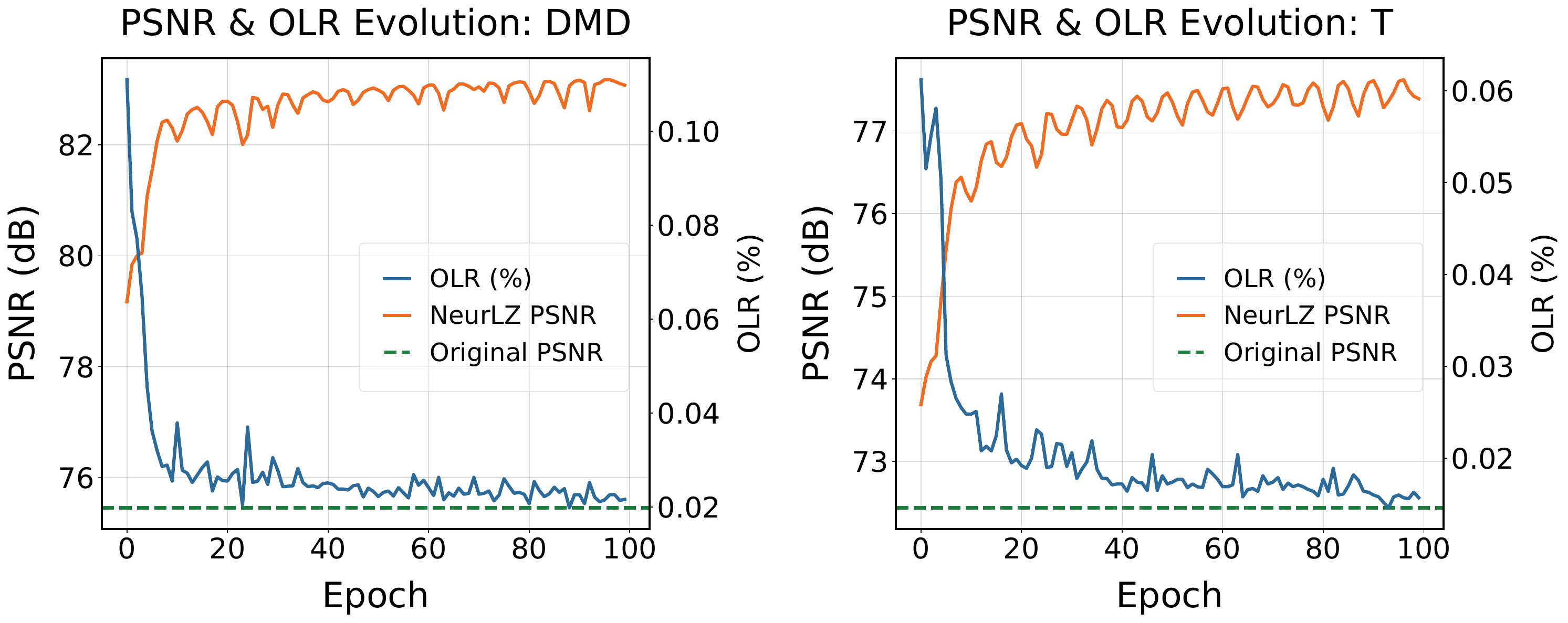}
  \vspace{-2\baselineskip}
  \caption{Training evolution of PSNR and outlier rate (OLR) (\%) for two fields in the Nyx, using \proj~ to enhance SZ3 with a 1E-3 relative error bound.}
  \label{fig:training-evolution}
  \vspace{-0.5\baselineskip}
\end{figure}

\FakeParagraph{Error validation} To validate the error distribution, we first analyze training dynamics. Figure \ref{fig:training-evolution} shows experiments on two Nyx fields, where \proj~enhances SZ3 with a 1E-3 bound. PSNR improves rapidly, then stabilizes around 100 epochs, and significantly exceeds the original. The Outlier Rate (OLR) (\%), representing the percentage of outliers, decreases sharply to 0.01\%, reducing the coordinate overhead for storing outlier locations. These improvements yield bit rate reductions of 66.4\% and 36.3\% at the same PSNR, respectively (Table \ref{tab:bitrate}).

\begin{figure}[t]
  \centering
  \includegraphics[width=\linewidth]{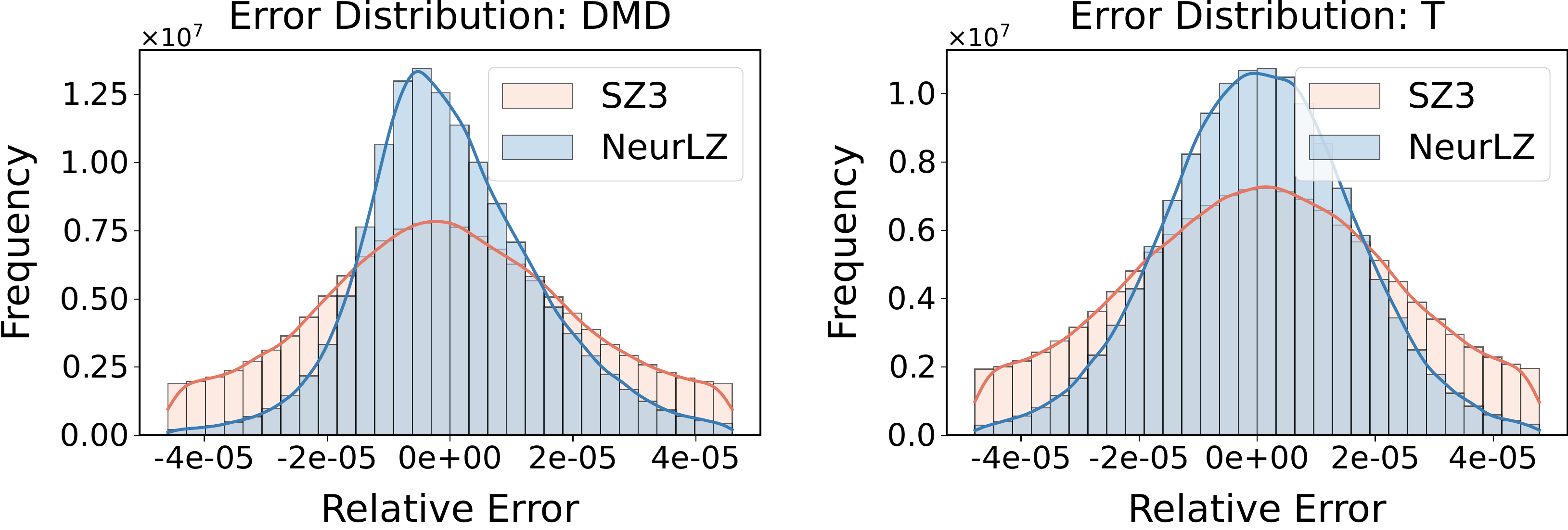}
  \vspace{-1.5\baselineskip}
  \caption{Error distribution for two Nyx fields under a 5E-5 relative error bound, comparing SZ3 decompressed and \proj-enhanced values to the original.}
  \Description{Histogram comparing error distributions for two Nyx fields under a 5E-5 relative error bound for SZ3 decompressed and \proj-enhanced values.}
  \label{fig:error-distribution}
  \vspace{-0.8\baselineskip}
\end{figure}

After learning, outlier coordinates are saved in the compressed archive. During decompression, outliers are replaced by in-bound decompressed values, ensuring that strict error bounds are maintained. Figure \ref{fig:error-distribution} clearly demonstrates the enhancement of SZ3 compression under a stricter targeted 5E-5 error bound, where \proj~confines all errors within this limit, with a more concentrated error distribution than that of SZ3.

\begin{figure}[t]
  \centering
  \includegraphics[width=\linewidth]{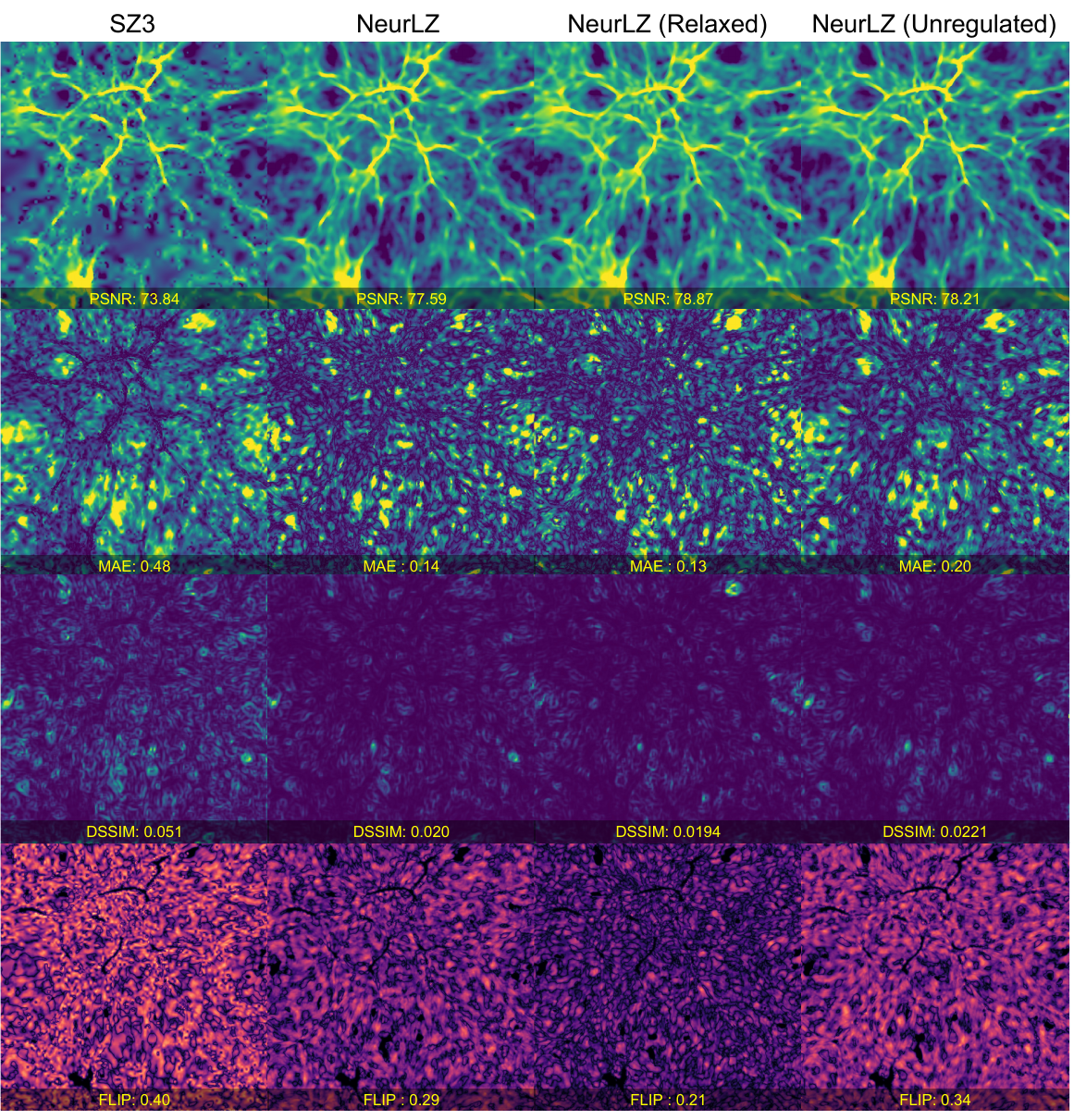}
  \vspace{-2\baselineskip}
  \caption{Comparison among SZ3, \proj, \proj~ (Relaxed), and \proj~ (Unregulated) on Nyx field T at the same bit rate. Higher PSNR and lower MAE, DSSIM, and FLIP are preferred. "Relaxed" relaxes strict error control with a regulated 2$\times$ error bound, while "Unregulated" has no enforced error bound.}
  \Description{Figure comparing SZ3, \proj, and \proj (Relaxed), and \proj (Unregulated) on a Nyx field T {\color{blue} FLIP error map and score: the smaller the better}}
  \label{fig:comparison}
  \vspace{-0.5\baselineskip}
\end{figure}

\FakeParagraph{Data visualization}
Figure \ref{fig:comparison} demonstrates the performance of SZ3, \proj~, \proj~ (Relaxed), and \proj~ (Unregulated) on Nyx field T at the same bit rate, evaluated using PSNR, MAE, DSSIM~\cite{Sara2019ImageQA}, and FLIP~\cite{flip}. While PSNR is commonly used to assess reconstruction quality, it does not fully account for human perceptual sensitivity to visual differences~\cite{Sara2019ImageQA, flip}. To address this, additional metrics are considered: MAE (Mean Absolute Error) quantifies pixel-wise reconstruction errors, DSSIM (Structural Dissimilarity Index) evaluates structural fidelity, and FLIP, originally proposed by NVIDIA, measures perceptual differences aligned with human vision. Higher PSNR values indicate better quality, and lower values for MAE, DSSIM, and FLIP indicate better performance overall.

Compared to SZ3, \proj~achieves significantly higher PSNR and lower MAE, DSSIM, and FLIP, highlighting its ability to produce reconstructions that are not only quantitatively superior but also more perceptually accurate. Between \proj~(Relaxed) and \proj~(Unregulated), the detailed analysis in Sec.~\ref{section_discussion_no_outlier} provides a further discussion on their trade-offs. However, as shown here, \proj~ (Relaxed) consistently delivers the highest PSNR and the lowest MAE, DSSIM, and FLIP among all methods, making it an appealing choice for users seeking objective accuracy and perceptual quality at the same bit rate.

\FakeParagraph{Bit rate vs. PSNR}  Bit rate reflects the average number of bits required to represent one data point, calculated from%
\[
  \operatorname{\it bit rate} =
  \frac{
    \texttt{size}(Z) + \operatorname{\it supplementary}
  }{
    \textit{num points of original data}
  },
\]
where $\texttt{size}(Z)$ is the size of compressed data $Z$.
The \textit{supplementary} includes storage for outlier coordinates and model parameters. Thus, the bit rate reflects the compressed data size and the storage costs of outliers and model parameters, providing a comprehensive measure of average storage cost.
Figure \ref{fig:bitrate} (left) presents the Bit rate vs. PSNR curves for SZ3, \sproj, and \proj~ across different Nyx fields, comparing their compression performance under varying error bounds. \sproj~ applies single-field learning, while \proj~ incorporates cross-field learning.

\proj~and \sproj~curves consistently shift to the upper right compared to SZ3 across all fields, indicating higher PSNR at the same bit rate or lower bit rate for the same PSNR. This improvement is particularly pronounced in fields like Dark Matter Density and Baryon Density.
Interestingly, \proj~does not always mirror \sproj's movement. For example, in the Baryon Density field, \proj's final point lies above and to the right of \sproj, achieving a lower bit rate and higher PSNR. This advantage stems from \proj's use of cross-field information, which reduces the proportion of outliers and thereby decreases overhead.

Beyond SZ3, \proj's impact on ZFP is also evaluated. Figure~\ref{fig:bitrate} (right) presents Bit rate vs. PSNR curves for ZFP, \sproj, and \proj~ across Nyx fields. For velocity fields, \sproj~shows no improvement over ZFP at higher bit rates, and \proj's gains diminish as the bit rate increases.
For fields such as Temperature, Dark Matter Density, and Baryon Density, \proj~consistently outperforms \sproj, and both significantly surpass the baseline ZFP. These results highlight \proj's ability to capture dependencies overlooked by ZFP, particularly through leveraging cross-field relationships. However, as ZFP converts data into a coefficient space, achieving further improvements may require unique architectural designs tailored to transform-based compressors. Nevertheless, the significant enhancements observed in these fields validate \proj's potential to improve compression efficiency in such frameworks.

\FakeParagraph{Bit rate reduction} To quantify \proj's improvements, three datasets (Table \ref{tab:dataset}) are tested with varying relative error bounds on SZ3 and ZFP. SZ3 uses bounds from 1E-2 to 1E-4, while ZFP uses 1E-1 to 1E-3 to maintain comparable PSNR ranges. ZFP's conservative error distribution results in higher PSNR under the same bound~\cite{7967203}, necessitating more relaxed bounds. Table \ref{tab:bitrate} presents the relative bit rate reduction (\%) at equal PSNR.

\proj~ demonstrates significant improvements with SZ3, particularly for the Nyx dataset. \Circled{1} For Baryon Density, the bit rate reduction reaches 91.3\% at a 1E-3 error bound, with both it and Dark Matter Density exceeding over 40\% reductions across all bounds. Temperature achieves 62.9\% at 1E-2, but reductions decrease for stricter bounds, except for Baryon Density, which improves from 50.1\% at 1E-2 to 69.0\% at 1E-4, suggesting \proj~ captures more relationships under stricter bounds.
\Circled{2}
For Miranda, most fields see ~20\% reductions at 1E-4, following the trend of lower reductions with stricter bounds. Although less pronounced than Nyx, the improvements remain consistent.
\Circled{3}
For Hurricane, reductions are weaker, under 10\% at 1E-4, with a maximum of 36.8\% for W at 1E-2. Hurricane's unique characteristics as a climate dataset may account for the smaller gains compared to Nyx and Miranda.

As for ZFP, \proj~ achieves the largest bit rate reductions of 81.8\%, 29.3\%, and 28.1\% for Nyx, Miranda, and Hurricane, respectively. While lower than SZ3's 91.3\%, 37.4\%, and 36.8\%, they remain significant. In Nyx, Temperature and Baryon Density see the largest reductions at 1E-2, not the loosest 1E-1 bound. Miranda shows stable reductions above 10\% across all tested bounds without SZ3's dynamic variations. For Hurricane, reductions are smallest (<5\%) at stricter bounds and largest (20\%) at looser bounds. These results collectively show \proj's effectiveness on ZFP, despite its transform-based nature posing more significant challenges than prediction-based SZ3.

\begin{table}[t!]
  \centering\footnotesize\sffamily
  \caption{{\proj}-achieved relative reduction (\%) in bit rate at equal PSNR across three datasets for SZ3 and ZFP; higher values indicate better overall performance.}
  \vspace{-0.8\baselineskip}
  \renewcommand{\arraystretch}{1.3}
  \setlength{\tabcolsep}{2.5pt}
  \resizebox{\linewidth}{!}{%
    \begin{tabular}{ | cl|ccccc|ccccc | }
    \hline
    \multicolumn{2}{|c|}{\textbf{Compressor}} & \multicolumn{5}{c|}{\textbf{SZ3}}                                                                          & \multicolumn{5}{c|}{\textbf{ZFP}}                                                                \\
    \hline
    \multicolumn{2}{|c|}{Error Bound}         & 1E-2                                                                                                       & 5E-3                              & 1E-3 & 5E-4 & 1E-4 & 1E-1 & 5E-2 & 1E-2 & 5E-3 & 1E-3        \\
    \hline
    \multirow{4}[2]{*}{\rotatebox{90}{\ Nyx}}
                                              & 1. \scriptsize Temperature                                                                                 & 62.9                              & 51.2 & 36.3 & 37.0 & 21.8 & 25.5 & 25.6 & 28.0 & 27.0 & 21.8 \\
                                              & 2. \scriptsize\raisebox{1pt}[0pt][0pt]{\begin{tabular}{@{}l@{}}Dark Matter\\[-1.5ex] Density\end{tabular}}
                                              & 89.1                                                                                                       & 86.4                              & 66.4 & 57.2 & 40.5 & 53.4 & 54.9 & 35.4 & 36.7 & 24.7        \\
                                              & 3. Velocity Y                                                                                              & 44.5                              & 41.5 & 36.5 & 24.0 & 1.17 & 28.6 & 29.5 & 21.6 & 15.1 & 3.67 \\
                                              & 4. \scriptsize\raisebox{2pt}[0pt][0pt]{\begin{tabular}{@{}l@{}}Baryon\\[-1.5ex] Density\end{tabular}}      & 50.1                              & 76.8 & 91.3 & 91.1 & 69.0 & 40.1 & 75.3 & 81.8 & 79.0 & 38.4 \\
    \hline
    \multirow{3}[2]{*}{\rotatebox{90}{\ \ Miranda}}
                                              & 1. Diffusivity                                                                                             & 37.4                              & 36.2 & 32.8 & 30.3 & 21.9 & 12.4 & 29.2 & 22.0 & 19.3 & 11.4 \\
                                              & 2. Velocity Z                                                                                              & 29.9                              & 28.1 & 28.3 & 25.1 & 23.0 & 10.6 & 26.1 & 26.0 & 23.9 & 14.8 \\
                                              & 3. Viscocity                                                                                               & 36.5                              & 36.0 & 31.6 & 30.7 & 20.8 & 11.8 & 29.3 & 21.8 & 18.0 & 13.6 \\
    \hline
    \multirow{3}[2]{*}{\rotatebox{90}{\ \ Hurricane}}
                                              & 1. Cloud                                                                                                   & 27.7                              & 16.9 & 4.54 & 5.07 & 4.34 & 15.3 & 13.7 & 8.79 & 6.30 & 4.43 \\
                                              & 2. Precip                                                                                                  & 24.2                              & 30.9 & 19.3 & 14.2 & 4.85 & 16.3 & 14.5 & 10.5 & 9.03 & 5.26 \\
                                              & 3. W                                                                                                       & 36.8                              & 30.6 & 24.9 & 18.1 & 6.50 & 28.1 & 25.3 & 16.7 & 13.3 & 6.46 \\
    \hline
  \end{tabular}%
  }
  \label{tab:bitrate}%
  \vspace{-1.6\baselineskip}
\end{table}%

\FakeParagraph{Scalability testing}
Table \ref{tab:scalability} demonstrates \proj's scalability on a Nyx field across dimensions and training epochs. Larger dimensions achieve higher bit rate reductions, up to 93.9\% for 2,048 at 10 epochs. This cross-epoch trend suggests our memorizer captures broader spatiotemporal and cross-field patterns compared to the conventional local feature search in SZ3~\cite{liang2022sz3}, as discussed in Sec.~\ref{section_discussion_Crossfield}. With larger data dimensions, our memorizer naturally attends to a wider data range in a single pass, offering a broader attention scope. This gives it a distinct advantage over SZ3, whose focus is constrained to local features.

For size 2,048, \proj~ achieves 80.0\% reduction in 1 epoch, with training time at 41.9\% of SZ3's compression time. Inference time remains constant across epochs, maintaining a 46.3\% decompression time. As dimensions grow, the reduced ratios highlight \proj's efficiency for large data, dropping from 95.7\% and 93.6\% for 512 to 41.9\% and 46.3\% for 2048, enabled by convolutional networks' ability to process large data regions in a single pass.

The training-time overhead compared to SZ3 is justified by 1) the long-term storage benefits~\cite{amazon_s3_glacier, azure_archive_storage} and 2) the future elimination by parallel optimization. \proj’s strong scalability creates opportunities to explore its performance on even larger datasets in the future.

\section{Discussions}

\begin{table}[t]
  \centering\footnotesize\sffamily
  \renewcommand{\arraystretch}{1.3}
  \caption{The {\proj}-achieved relative reduction (\%) in bit rate at equal PSNR on a Nyx field across varying dimensions and training epochs for SZ3.}
    \vspace{-\baselineskip}
   \begin{tabular}{ |l|l|r|r|r|r|r|}
    \hline
    \bfseries Size                          &
    {\textbf{Metrics ($\downarrow$)\,/\,Epochs ($\rightarrow$)}}
                              &
    \bfseries 1               &
    \bfseries 2               &
    \bfseries 5               &
    \bfseries 10
    \\
    \hline
    \multirow{3}{*}{$512^3$}  & Rate Reduction (\%)    & 8.3              & 41.0              & 51.5              & 72.9              \\
                              & Train/Comp. Time (\%)  & 95.7             & 168.$\phantom{0}$ & 539.$\phantom{0}$ & 930.$\phantom{0}$ \\
                              & Inf./Decomp. Time (\%) & 93.6             & 93.6              & 93.6              & 93.6              \\
    \hline
    \multirow{3}{*}{$1024^3$} & Rate Reduction (\%)    & 45.4             & 61.3              & 75.9              & 82.6              \\
                              & Train/Comp. Time (\%)  & 48.$\phantom{0}$ & 104.$\phantom{0}$ & 240.$\phantom{0}$ & 475.$\phantom{0}$ \\
                              & Inf./Decomp. Time (\%) & 59.2             & 59.2              & 59.2              & 59.2              \\
    \hline
    \multirow{3}{*}{$2048^3$} & Rate Reduction (\%)    & 80.0             & 84.2              & 88.6              & 93.9              \\
                              & Train/Comp. Time (\%)  & 41.9             & 83.6              & 209.$\phantom{0}$ & 418.$\phantom{0}$ \\
                              & Inf./Decomp. Time (\%) & 46.3             & 46.3              & 46.3              & 46.3              \\
    \hline
  \end{tabular}%

  \label{tab:scalability}%
  \vspace{-1.2\baselineskip}
\end{table}%

\label{section_discussion}
\subsection{Doubling the Error Bound}
\label{section_discussion_no_outlier}

\begin{figure}[t!]
  \centering
  \includegraphics[width=\linewidth]{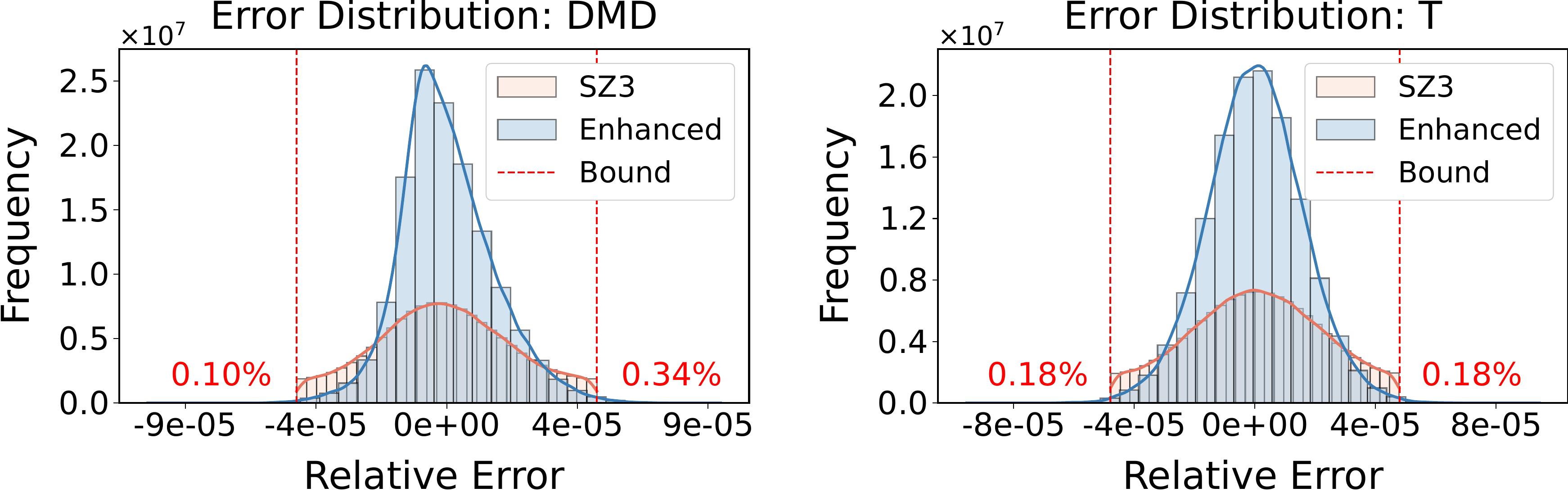}
  \vspace{-\baselineskip}
  \caption{Error distribution for two Nyx fields---SZ3 decompressed vs. \proj-initial-enhanced values. Outlier rates are 0.10\%–0.34\% for DMD and 0.18\% for T.}
  \Description{Error distribution comparison for two Nyx fields: DMD and T.}
  \label{fig:bad-error-distribution}
  \vspace{-0.5\baselineskip}
\end{figure}

The central question surrounding error regulation is
\begin{formal}
  {\it whether doubling the error bound can eliminate the need for storing outlier}.
\end{formal}
As discussed in Section \ref{section_error-bound}, \proj~ achieves  strict $1\times$ error bounds by storing outlier coordinates. Figure \ref{fig:bad-error-distribution} shows the error distribution, along with the percentage of outliers under a 5E-5 bound for two fields, comparing SZ3 decompressed values with \proj~ initial enhanced values. To maintain strict error control, final enhanced values are obtained by replacing outliers with decompressed values (Figure \ref{fig:outlier}), which necessitates storing outlier coordinates. This results in bit rate savings of 20.9\% and 15.6\% for these two fields, respectively.

By skipping the storage of outlier coordinates, however, compression efficiency improves significantly, with bit rate savings increasing to 37.6\% and 25.7\%, providing over 10\% additional storage reduction in both cases.
More importantly, relaxing strict error control does not noticeably compromise reconstruction quality. As discussed in Sec.~\ref{section_error-bound}, the regulation mechanism in \proj~ ensures that errors remain within a $2\times$ bound. Figure \ref{fig:comparison} further illustrates that \proj~ (Relaxed) leverages this mechanism to achieve higher reconstruction quality than strict error control (\proj) at the same bit rate.

Regarding the importance of regulation, we test {\proj} (Unregulated), which removes both strict error control and regulation mechanisms. While it achieves a PSNR of 78.21 (even higher than \proj), its lack of constraints results in significantly worse MAE, DSSIM, and FLIP scores. This aligns with the observations in Figure \ref{fig:2xbound-evolution}, where a controlled experiment demonstrated that Unregulated and Regulated approaches achieve similar final PSNR values. However, the absence of regulation in the Unregulated approach leads to substantially higher overall error magnitudes. These findings highlight that our regulation is essential for providing a flexible and practical trade-off between compression efficiency and reconstruction quality.

\subsection{Uncovering Cross-Field Contributions}
\label{section_discussion_Crossfield}
The central question surrounding cross-field learning is
\begin{formal}
  \it how different fields contribute to predictions.
\end{formal}
Figure \ref{fig:effect} shows that cross-field learning achieves higher PSNR than single-field learning. Information from multiple fields enables more accurate predictions, enhancing PSNR. Understanding how different fields contribute to this improvement is crucial -- how much does each field contribute? Are all fields equally helpful? We analyze the Temperature field using SZ3 with a 5E-3 bound. Incorporating two other Nyx fields into cross-field learning, we train the model for 100 epochs and use Captum with integrated gradients for attribution analysis~\cite{kokhlikyan2020captum, sundararajan2017axiomatic}.

\begin{figure}[t!]
  \centering
  \includegraphics[width=\linewidth]{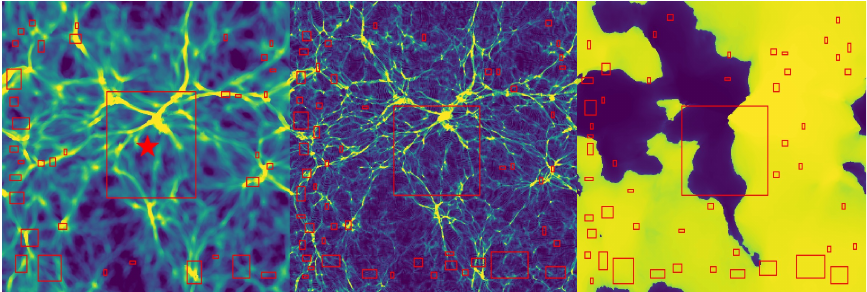}
  \vspace{-\baselineskip}
  \caption{Significant attribution regions (red rectangles) are overlaid on slices from three Nyx fields—T, DMD, and Vx—highlighting the areas that contribute to the prediction at the center of the T field (red star).}
  \Description{The figure highlights significant attribution regions contributing to T field prediction.}
  \label{fig:attr}
  \vspace{-0.5\baselineskip}
\end{figure}

Using decompressed slices ($512 \times 512$) from these three fields, the model predicts the residual error for Temperature. Attribution analysis targets the datum at position (256, 256) in the Temperature slice. Captum computes scores quantifying each field's contribution to this prediction. Figure \ref{fig:attr} shows the results: the left slice displays the target marked by a red star, while significant attribution regions are highlighted with red rectangles across all slices for clarity.

Each field shows a prominent red rectangle around (256, 256), corresponding to the predicted datum, clearly highlighting \proj's identification of local patterns near the target across fields. Additionally, smaller rectangles distributed across the fields indicate global patterns contributing to the prediction. The variation in local and global patterns among fields suggests differing levels of contributions from each field. This visualization effectively demonstrates that \proj~ leverages adaptable, learnable cross-field patterns, enhancing the reconstruction quality.

\subsection{Analyzing Performance Limitations}
\label{section_discussion_conflict}

\begin{figure}[t!]
  \centering
  \includegraphics[width=\linewidth]{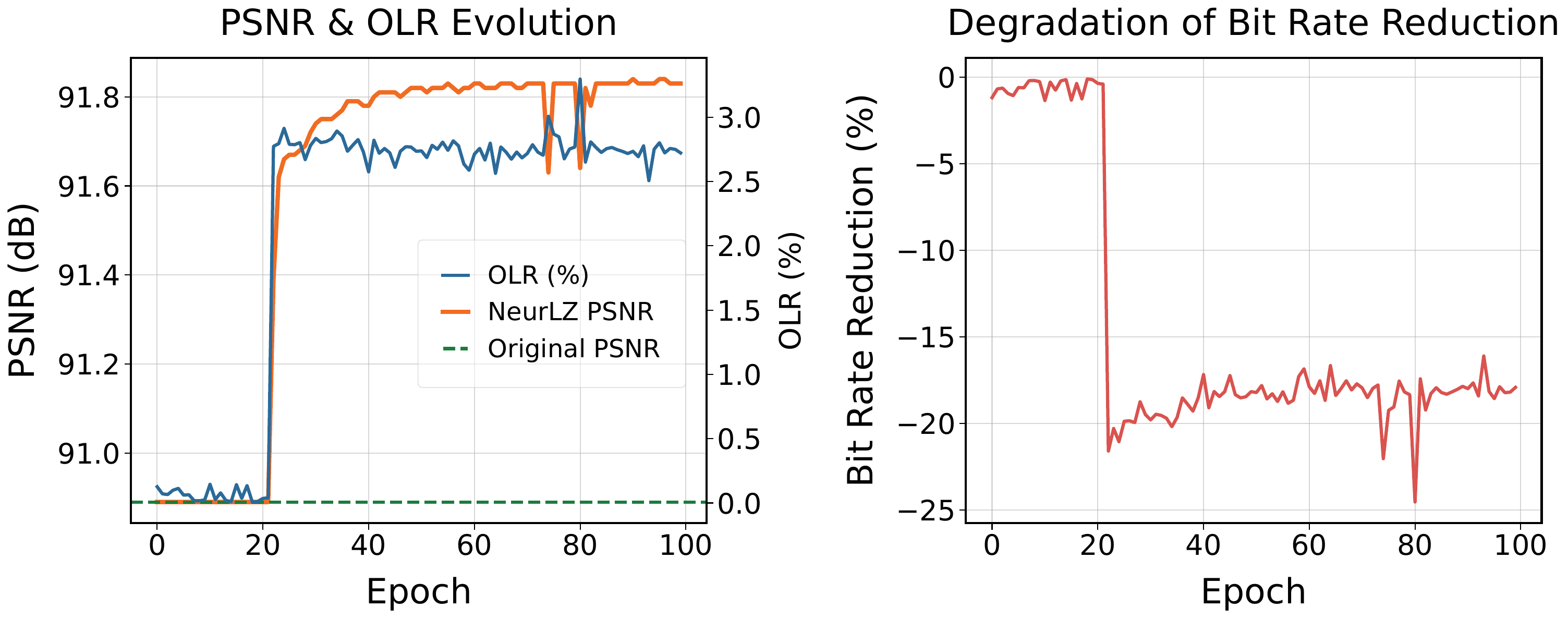}
  \vspace{-\baselineskip}
  \caption{Training evolution of PSNR and outlier rate (OLR, \%) for one Nyx field. The sharp PSNR rise, high OLR, and eventual degradation of bit rate reduction indicate challenges for \proj~ in balancing PSNR improvement and maintaining a low OLR.}
  \Description{Line charts showing the training evolution of PSNR and outlier rate (OLR) for the Vx field in Nyx. Results highlight a sharp PSNR rise, high OLR, and reduced bit rate reduction, indicating challenges for the skipping DNN in maintaining a balance.}
  \label{fig:bad-training-evolution}
  \vspace{-1\baselineskip}
\end{figure}

The central question surrounding performance limitations is
\begin{formal}
  \it how conflicts affect \proj~at stricter error bounds.
\end{formal}
As discussed in Section \ref{section_experimental-results}, stricter error bounds reduce \proj's effectiveness. For example, in the Nyx Velocity Y field with a 1E-4 error bound, only a 1.17\% bit rate reduction is achieved by \proj~ (Table \ref{tab:bitrate}), and with smaller bounds like 5E-5, no improvement may occur.
Figure \ref{fig:bad-training-evolution} shows the PSNR and outlier rate (OLR) evolution during training for Velocity Y with a 5E-5 error bound. PSNR plateaus for approximately 20 epochs, then briefly rises before stagnating, while OLR increases alongside PSNR, leading to a negative bit rate reduction, indicating a potentially detrimental effect. This contrasts with cases in Figure \ref{fig:training-evolution}, where PSNR rises, OLR decreases, and bit rate reductions of 66.4\% and 36.3\% are achieved. These results suggest that stricter error bounds leave \proj~ with less room for optimization, raising the question: why is there a performance limitation with stricter error bounds?

We hypothesize that sample conflict introduces noise and further hinders \proj~performance at stricter error bounds. This occurs when input data $x_1$ and $x_2$ are very similar or identical ($x_1 \approx x_2$ or $x_1 = x_2$), but their desired outputs $y_1$ and $y_2$ differ significantly ($y_1 \neq y_2$). In such cases, the model struggles to learn effectively, as it must map nearly identical inputs to distinct outputs ($f(x_1) = y_1$, $f(x_2) = y_2$). Instead, the model naturally tends to map similar inputs to similar outputs ($f(x_1) \approx f(x_2)$), conflicting with the objective of $y_1 \neq y_2$. This inherent challenge makes it difficult to fit the training data properly.

In our case, $x$ represents the decompressed slice, and $y$ represents the target field's residual slice. To illustrate, we analyze two fields from the Nyx dataset—Temperature and Velocity Y—under SZ3, focusing on single-field learning. This approach can naturally be extended to cross-field learning scenarios. These fields were selected due to their contrasting behaviors: Temperature shows notable improvements even under stricter error bounds, while Velocity Y lacks similar gains. Each decompressed slice, with a shape of 512 $\times$ 512, corresponds to an input $x$, while each residual slice, also with a shape of 512 $\times$ 512, corresponds to the output $y$. We can compute the similarity between different $x$ slices and between different $y$ slices using the absolute value of cosine similarity as the metric. If the similarity between two decompressed slices, $x_1$ and $x_2$, exceeds 0.95 while the similarity between their corresponding residual slices, $y_1$ and $y_2$, is less than 0.05, we consider $(x_1, y_1)$ and $(x_2, y_2)$ to be in a conflict. Otherwise, they are not considered as being in conflict under this criterion.

\begin{figure}[t!]
  \centering
  \includegraphics[width=\linewidth]{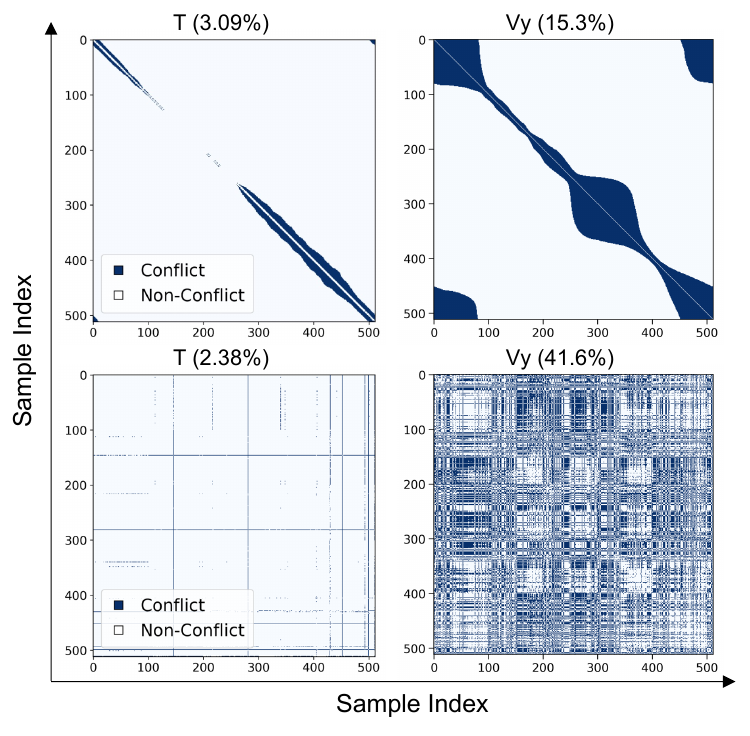}
  \vspace{-2.1\baselineskip}
  \caption{Sample (top) and gradient (bottom) conflict matrices for two Nyx fields, each with 512 samples. The matrices show conflicts across 262,144 pairs, where white (0) indicates no conflict, dark blue (1) indicates conflict, with titles showing conflict proportions.}
  \Description{Conflict matrices for two Nyx fields, showing sample conflicts (top) and gradient conflicts (bottom) across 262,144 pairs.}
  \label{fig:conflict}
  \vspace{-0.2\baselineskip}
\end{figure}

The upper part of Figure \ref{fig:conflict} shows the sample conflict adjacency matrix. Since there are 512 $(x_i, y_i)$ pairs, the matrix has a shape of 512 $\times$ 512, representing a total of 262,144 pairwise comparisons for conflict calculation. The matrices depict the occurrence of conflicts between these pairs: white areas (value 0) indicate non-conflicting pairs, while dark blue areas (value 1) represent conflicting pairs. The subplot titles display the proportion of conflicts, quantifying the percentage of pairs exhibiting conflicts. In the left figure for Temperature, fewer conflicts are observed, with only 3.09\% of the total pairs showing conflicts, compared to 15.3\% for Velocity Y. This aligns with the fact that Velocity Y is more challenging to enhance. Interestingly, for both cases, most conflicts occur between neighboring slices, such as $(x_i, y_i)$ and $(x_{i+1}, y_{i+1})$, which is visible as a dark blue region along the line $y = x$. This may be due to the decompressed values being smoother than the residual values, as the residuals tend to be more random \cite{liang2022sz3}. As a result, similar $x_i$ and $x_{i+1}$ values but different $y_i$ and $y_{i+1}$ values lead to conflicts.

Conflicts among samples can cause conflicting gradient updates during training. Models were trained for 100 epochs, and gradients for all 3,000 parameters were extracted per sample, forming gradient vectors of size 3,000 $\times$ 1. Gradient conflict adjacency matrices were calculated based on these gradient vectors. The lower part of Figure \ref{fig:conflict} shows these matrices for Temperature and Velocity Y. Temperature has a low conflict rate of 2.38\%, indicating similar gradient directions and consistent updates. In contrast, Velocity Y shows a high conflict rate of 41.6\%, with samples disagreeing on update directions.
These differences align with the training dynamics shown in Figures \ref{fig:training-evolution} and \ref{fig:bad-training-evolution}. For Temperature, consistent gradients result in smooth and steady PSNR increases, and OLR decreases. For Velocity Y, conflicting gradients lead to unstable PSNR and OLR increases, severely hindering optimization. Overall, sample and gradient conflicts inherently limit model performance, causing reduced PSNR gains and higher OLR, leading to diminished or even negative overall performance improvements.

\section{Conclusion}
In this paper, we propose \proj, an online neural learning-based method designed to enhance lossy compression quality in handling large-scale scientific data, fully leveraging the power of DNNs.
\proj~ dynamically trains multiple lightweight skipping DNN models during compression for specific data blocks, adapting efficiently to residual errors and evolving data characteristics.
Evaluations on diverse datasets such as Nyx, Miranda, and Hurricane show \proj's superior performance, achieving up to 94\% reduction in bit rate while maintaining equivalent levels of data distortion and significantly improving upon existing state-of-the-art conventional compression methods.

\clearpage
\bibliographystyle{ACM-Reference-Format}
\bibliography{ref}
\end{document}